\documentclass[a4paper,11pt]{article}

\usepackage{jcappub} 

\usepackage[T1]{fontenc} 
\usepackage{graphicx}
\usepackage{subfigure}
\usepackage{epstopdf}

\usepackage{verbatim}
\usepackage{ulem}
\usepackage{color}
\usepackage{float}

\usepackage{mathrsfs}
\usepackage{amssymb}
\usepackage{amsmath}
\usepackage{amsfonts}
\usepackage{amsthm}
\usepackage{psfrag}
\usepackage{epsfig}
\usepackage{bm}

\usepackage{lscape}
\usepackage{comment}
\usepackage{multirow}

\title{\boldmath Optimisation of the Population Monte Carlo algorithm: Application to constraining isocurvature models with cosmic microwave background data}

\author[1]{D. Moodley}
\author[1]{and K. Moodley}

\affiliation[1]{Astrophysics and Cosmology Research Unit \& School of Mathematics, Statistics and Computer Science,
  University of KwaZulu-Natal, Durban, 4041, South Africa}

\emailAdd{darell.ukzn@gmail.com}
\emailAdd{moodleyk41@ukzn.ac.za}

\newcommand{\Eff}{E}

\abstract{We optimise the parameters of the Population Monte Carlo algorithm using numerical simulations. The optimisation is based on an efficiency statistic related to the number of samples evaluated prior to convergence, and is applied to a D-dimensional Gaussian distribution to derive optimal scaling laws for the algorithm parameters. More complex distributions such as the banana and
bimodal distributions are also studied. We apply these results to a cosmological parameter estimation problem that uses CMB anisotropy data from the WMAP nine-year release to constrain a six parameter adiabatic model and a fifteen parameter admixture model, consisting
of correlated adiabatic and isocurvature perturbations. In the case of the adiabatic model and the admixture model we find respective degradation factors of three and twenty, relative to the optimal Gaussian case, due to degeneracies in the underlying parameter space. The WMAP nine-year data constrain the admixture model to have an isocurvature fraction of at most $36.3 \pm 2.8$ percent.}

\begin{document}
\maketitle
\flushbottom

\section{Introduction}

\noindent
There has been a dramatic increase in both the quantity and quality of cosmological data from various observational probes. These observational data have improved our understanding of the universe by awarding merit to theoretical models that fit the data well and providing estimates for parameters that these models depend on. Most notably, the cosmic microwave background anisotropy \citep{Runyan2003ApJS_149_265R, Kosowsky2003NewAR..47..939K, Tristram2004ASSL..301...97T, Ade2014ApJ_792_62A, MacTavish2006ApJ...647..799M, Padin2001ApJ...549L...1P, Mather1990ApJ...354L..37M, Rabii2006RScI...77g1101R, Bouchet2014arXiv1405_0439B, Miller2002ApJS..140..115M, Ruhl2004SPIE.5498...11R, Bennett2003ApJ_583_1B, Barkats2003NewAR..47.1077B, Leitch2002ApJ...568...28L, Johnson2003NewAR..47.1067J, Wollack1997ApJ...476..440W, Ade2008ApJ_674_22A, QUIET2012arXiv1207_5562Q, Watson2003MNRAS.341.1057W} has enabled precision parameter estimation at the percent level \citep{Planck2015arXiv150202114P, Planck2015arXiv150201589P, Bennett2013ApJS..208...20B}. Amongst various parameter estimation methods, the Bayesian sampling approach (\cite{Bayes63}; also see \cite{Lindley2005, sivia2006data,robert2007bayesian}, for example) is now the most commonly used in cosmology. Parameter estimation in the Bayesian framework is accomplished by estimating the posterior probability distribution, which usually requires evaluating multidimensional integrals. Evaluating these integrals can be achieved with the aid of efficient sampling algorithms.\\

\noindent
Sampling algorithms provide a sequence of samples which are approximated from the posterior distribution when direct sampling is difficult. This chain sequence, or sample set, can then be used to approximate quantities with respect to the posterior distribution. Sampling algorithms such as \textit{nested sampling} \citep{Skilling06} are widely used in cosmological parameter estimation \citep{Mukherjee06, Parkinson06, Nikolic09, Valiviita09} through publicly available software packages such as CosmoNest \citep{Mukherjee06, Parkinson06} and MultiNest \citep{Feroz09}. The most widely used sampling algorithm is \textit{Markov Chain Monte Carlo} (MCMC) \citep{Metropolis1953} with a publicly available package called COSMOMC \citep{2002PhRvD..66j3511L}.\\

\noindent
This paper focuses on a hybrid \textit{Adaptive Importance Sampling} algorithm \citep{Oh92} called \textit{Population Monte Carlo} (PMC) \citep{Cappe2004}, which has already been applied to the field of cosmology \citep{Wraith09, Dietrich2010MNRAS, Kilbinger10} with a publicly available code called CosmoPMC \citep{Kilbinger2011CosmoPMC}. PMC is iterative in design and it generates independent sample sets at each iteration which can be used for parameter estimation. This enables parallelisation of a single sampling run thereby distributing the likelihood evaluation for these models and reducing the computational time. The algorithm has its own convergence criterion enabling the sampler to automatically stop when this criterion has been satisfied.\\

\noindent
The PMC sampler runs independently of the user once the input sampling function and parameters are initialised, unlike MCMC which usually requires updating of the input covariance matrix. These inputs affect the performance of the sampler, measured in terms of its computational expense. With limited computational resources and the need for more feasible computational times, it is necessary to optimise PMC with regards to its algorithm parameters, the study of which is lacking in the literature. In this paper we find the optimal PMC algorithm parameters by optimising a statistic that measures the performance of the PMC algorithm. A similar study for MCMC was undertaken by \cite{Dunkley04}.\\

\noindent
We then apply the results of our optimisation study to a cosmological problem, specifically determining the relative contribution of adiabatic and isocurvature modes permitted by WMAP 9-year data to the cosmic microwave background spectrum. Constraining models with mixtures of adiabatic and isocurvature perturbations becomes more difficult since the complexity of the distributions increases as more isocurvature modes are added to the model. The higher dimensional parameter space introduces flat directions that are challenging for any sampler. It is therefore necessary to establish a reliable and effective way of sampling these distributions. This paper addresses the problem using the optimised PMC algorithm.\\

\noindent
We present constraints on models with correlated adiabatic and isocurvature modes derived from our study. Constraints on an adiabatic mode correlated with one isocurvature mode have been presented in \cite{Vita2012ApJ_753_151V}, \cite{Savelainen2013PhRvD_88f3010S} and \cite{Planck2015arXiv150202114P} using recent data from the WMAP 7-year and 9-year experiments and the Planck experiment, respectively. The new constraints on admixtures of the adiabatic mode with three isocurvature modes presented here are the most recent results for this case. A follow-up paper will extend this study to the latest Planck data \citep{Planck2015arXiv150702704P}.
The paper is structured as follows. Section \ref{sec:PMC} briefly describes Bayesian parameter estimation and the PMC algorithm. Section \ref{sec:Converge_Effic_Sim} describes the statistic used to assess convergence of a PMC chain and compares it to a convergence statistic commonly used in MCMC studies. We also describe a measure of efficiency and the simulations used to optimise this measure. Section \ref{sec:OptimisePMC} contains results of our study to optimise the PMC algorithm for a Gaussian target distribution using simulations. Section \ref{sec:complextarg} assesses the degradation of efficiency for non-ideal target distributions such as the banana shaped and bimodal distributions. In Section \ref{sec:CosmoProblem} we apply our findings from Section \ref{sec:OptimisePMC} and \ref{sec:complextarg} to the cosmological parameter estimation problem of constraining a mixed adiabatic and isocurvature perturbation model. Section \ref{sec:conclusion} presents our concluding remarks.

\section{The PMC algorithm}
\label{sec:PMC}

The posterior probability distribution, $\pi$, of a set of parameters $x$, given a data set, $D$ can be inferred using \textit{Bayes' theorem}:
\begin{equation}
 \pi(x|D) = \frac{P(D|x)P(x)}{P(D)} ,
\label{eqn:Bayestheorem}
\end{equation}
where $P(D|x)$ and $P(x)$ are known as the \textit{likelihood} and \textit{prior} probability respectively. The denominator, $P(D)$, referred to as the evidence is relevant in model selection but not parameter estimation, therefore we do not consider it in this study. We omit the dependence on data for convenience and write $\pi(x)$ instead throughout this paper. We most often lack an analytical expression for the posterior therefore we rely on using a sample from this distribution to evaluate any integrals related to $\pi(x)$.\\

PMC is an iterative sampling algorithm that aims to improve estimates of parameters using the sequence of samples generated after each iteration. We independently draw a sample $(x_1,x_2,\cdots,x_{N_s})$ from $q$, where $q$ is known as the \textit{importance} function and $N_{s}$ is the number of samples evaluated per iteration. Hence, for some function $f$ we have a convergent estimator,
\begin{equation}
\left< f_{N}\right> \approx \frac{1}{N_s}\sum_{i=1}^{N_s}f(x_n)\bar{w}_n ,
\end{equation}
with normalised importance weights, 
\begin{equation}
\bar{w}_n = \frac{w_n}{\sum_{i=1}^{N_s}w_i },
\end{equation}
and $w_n = \frac{\pi(x_n)}{q(x_n)}$. The \textit{Kullback-Liebler} divergence \citep{Kullback1951}, denoted $K$, based on the closeness between $q$ and $\pi(x)$, is used to set the convergence criterion for the sampler. A linear sum of a number of mixture densities or components for $q$ are suggested by \cite{Cappe07} instead of an explicit form. That is,
\begin{equation}
q^t(x^t,\alpha ^t,\theta ^t)=\sum_{c=1}^{N_c}\alpha_{c}^t \phi_{c} (x^t,\theta_{c}^t),
\label{eqn:mixturedensities}
\end{equation}
where $\boldsymbol{\alpha}^t =(\alpha_{1}^t,\alpha_{2}^t,\cdots,\alpha_{N_c}^t)$ are the component weights associated with the sample size chosen from each respective component, $\phi_c$, for $N_c$ components, and $\theta^t_c$ are parameters of the component distributions. Each component has distribution parameters $\theta$. It is suggested \citep{Wraith09} that the $\phi_c$ distributions chosen are either multivariate Gaussian or Student-t distributions with the latter chosen in the case that $\pi(x)$ is suspected to have heavy tails. In this paper we find t-distributions to be less efficient in general than Gaussian distributions, which proved to be sufficient for our studies. We therefore use Gaussian mixture densities of dimension $D$ specified by a covariance matrix $\Sigma_{c}$ and mean $\mu_{c}$. We refer the reader to \cite{Cappe2004} for details of the updating procedure.

\section{Assessing convergence and efficiency of the PMC algorithm}
\label{sec:Converge_Effic_Sim}
In this section we establish a convergence statistic for PMC which we compare to a statistic used in MCMC. We also define a measure of efficiency for the PMC sampler, which is based on the total number of samples evaluated until convergence is reached.

\subsection{A sufficient convergence statistic}
\label{sec:ESS_vs_r}
Following the work of \cite{Wraith09}, we use an estimator of the Kullback-Liebler divergence, $K,$ at each iteration $t$,
\begin{equation}
\mbox{ESS}^t= \frac{1}{N_s}\left( \sum_{n=1}^{N_s} (\bar{w}_{n}^t)^2 \right)^{-1},
\label{eqn:ESS}
\end{equation}
which is called the normalised \textit{Effective Sample Size}, with $\mbox{ESS} \geq 95\%$ indicating that convergence has been achieved \citep{Wraith09}, from our simulations described below.\\

We compare this to another commonly used convergence statistic introduced by \cite{Dunkley04},
\begin{equation}
 r = \frac{\sigma_{\bar{x}}^2}{\sigma_{\pi}^2},
\label{eqn:r_stat}
\end{equation}
where $\sigma_{\bar{x}}^2$ is the variance from the sample mean and $\sigma_{\pi}^2$ is the variance of the target distribution. Convergence is achieved when $r < 0.01$, which indicates that the mean of the sample is close enough to the target mean.\\

We first conduct a simulation to compare the effectiveness of both these statistics by sampling from a one-dimensional Gaussian target distribution. We use 100 realisations over a wide range of sample sizes. We find that for small sample sizes, $N_s < 50$, the statistic $r$ is far stricter than ESS since a greater number of iterations is required to satisfy the $r$ convergence criteria. However for large enough sample sizes, $N_s > 200$, our results indicate that ESS is just as strict a constraint as $r$.\\

Our simulations thus suggest that when using the ESS statistic, it is advisable to choose the sample size large enough. In practice we find that it is non-optimal to use small sample sizes, so this criterion is automatically satisfied. We investigate the choice of an optimal sample size in Section \ref{sec:optimNs} but first we derive a quantity to measure the performance of PMC.

\subsection{Defining efficiency}
\label{sec:DefEfficiency}
An efficient sampler achieves convergence with minimal computational time. The computational cost is related to the number of evaluations at iteration $t$ by
\begin{equation}
N_{cost,t}=N_{c}N_s,
\label{eqn:Neff}
\end{equation}
assuming the number of components remains constant at each iteration. After $T$ iterations we have the total number of evaluations,
\begin{equation}
N_{Total}	=  \sum_{t=1}^{T} N_{cost,t} = TN_{c}N_{s},
\label{eqn:Ntotal}
\end{equation}
where $N_{Total}$ is the total number of evaluations. In the case that $N_c$ varies we choose to sum up the cost after each iteration. Note that we are not restricted to a constant sample size at each iteration \citep{Chopin2002}, therefore equation (\ref{eqn:Neff}) can be modified as necessary. For the purpose of this study however, we keep the sample size constant at each iteration. The efficiency, $\Eff$, is then measured by,
\begin{equation}
\Eff =\left( N_{Total} \right)^{-1}.
\label{eqn:efficiency}
\end{equation}

\subsection{Simulations}
\label{sec:SimMethod}
We use a simulation with known target distributions to study how $\Eff$ depends on the PMC algorithm parameters, which in turn allows us to determine the optimal choices of these parameters. As seen previously we are required to initially set the number of components, $N_c$, and the number of samples per iteration, $N_s$. In addition we need to initialise the mixture densities, which we choose to be Gaussian distributions with mean, $\mu_c,$ and diagonal covariance, $\Sigma_{c}=\mbox{diag}(\sigma_{c}^2,\cdots,\sigma_{c}^2)$. We choose to sample $\sigma_{c}$ and $\mu_c$ from uniform distributions as follows:
\begin{equation}
 \sigma_{c} \sim \mbox{UNIFORM}(\sigma_{0}-\sigma_{1},\sigma_{0}+\sigma_{1}),
\end{equation}
with $\sigma_0$ and $\sigma_1$ representing the center and width of the uniform distribution, respectively, and similarly for the mean,
\begin{equation}
 \mu_{c} \sim \mbox{UNIFORM}(\mu_{0}-\mu_{1},\mu_{0}+\mu_{1}).
\end{equation}
This brings the number of PMC algorithm parameters to six, namely, $N_s, N_c, \sigma_0, \sigma_1, \mu_0,$ and $\mu_1$.\\

The simulation is conducted over a specific range of values for the PMC algorithm parameters, creating a discrete space over which $\Eff$ is measured. We use 30 realisations at each point in parameter space to reduce noise in the simulated data. To obtain the relationship of $\Eff$ against each PMC algorithm parameter, we average $\Eff$ over the remaining five parameters. This method enables us to determine the optimal values for each of the PMC algorithm parameters.

\section{Optimisation of the PMC parameters: Gaussian target distribution}
\label{sec:OptimisePMC}

In this section we consider the case of an idealised target distribution, namely a multivariate Gaussian distribution,
\begin{equation}
\pi(\mathbf{x}) \sim N(\mathbf{\mu},I), 
\end{equation}
with $I$ the identity matrix and a mean, $\mathbf{\mu},$ of unity in all dimensions. The Gaussian distribution is commonly found in cosmological datasets \citep{Kilbinger10, Wraith09}, and is also a canonical distribution from which extensions can be made to more complicated distributions. We consider more complex distributions in the next section.
Simulations are run for different initial values of the PMC algorithm parameters until they converge to the target. Since the mixture densities and the target distribution have the same functional form, the PMC algorithm is able to match $q$ very closely with $\pi.$ This allows us to impose a stricter requirement of $\mbox{ESS} \geq 98\%$ as our convergence criterion.
Averaging over many of these simulations allows us to explore how efficiency changes across the underlying parameter space. 
We are also interested in how the efficiency of the PMC algorithm scales with dimension therefore we run the above simulations in various dimensions.

\subsection{Mixture density parameters}
\label{sec:mixturedensities}
\begin{figure}
     \begin{center}
        
        \subfigure{
            \includegraphics[width=70mm]{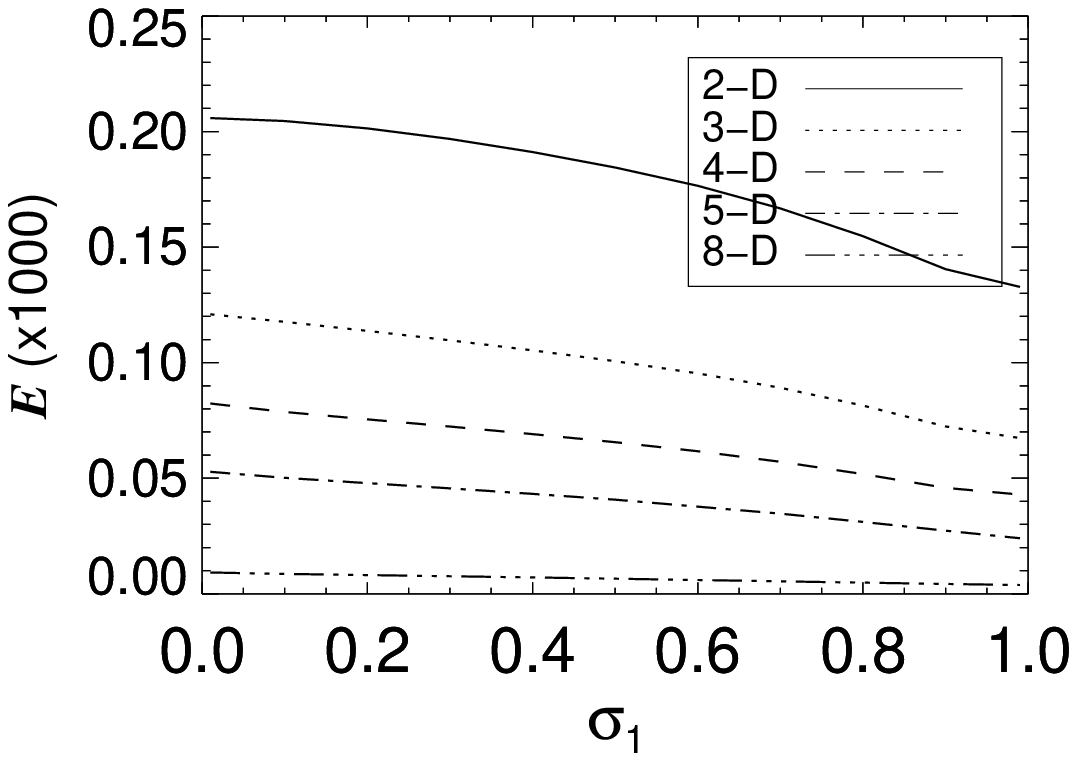}\hspace{-0.5cm}
        }
        \subfigure{
           \includegraphics[width=70mm]{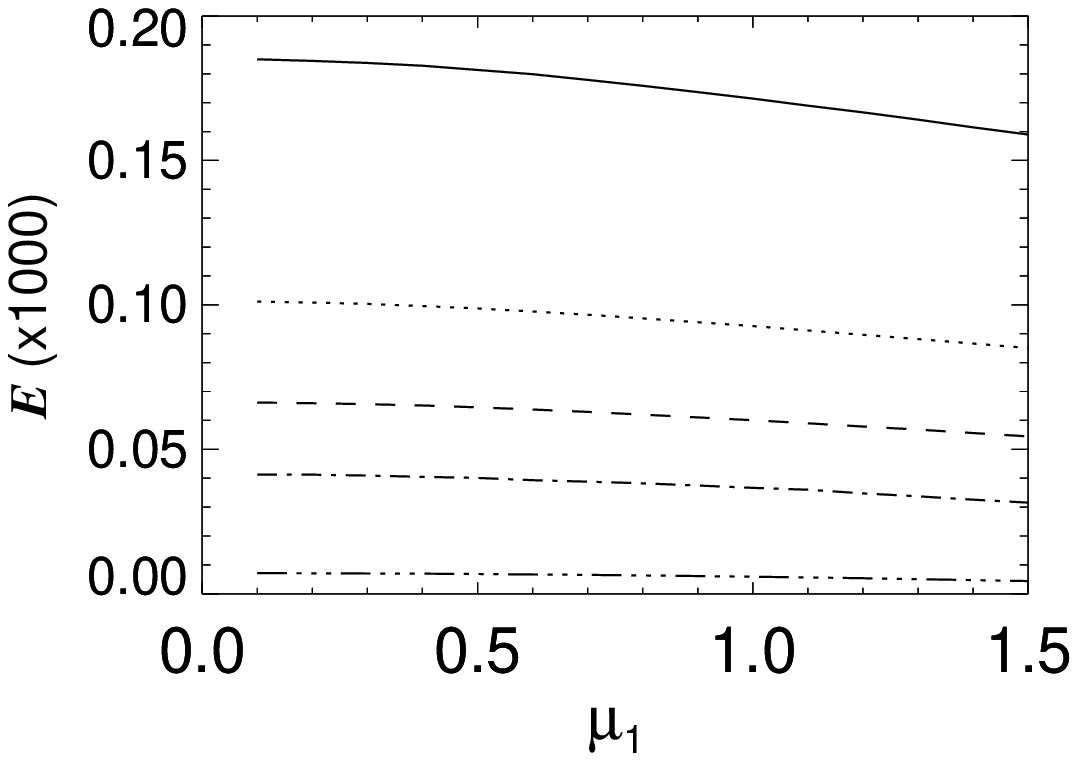}
        }\\ 
        \vspace{-0.5cm}
        \subfigure{
            \includegraphics[width=70mm]{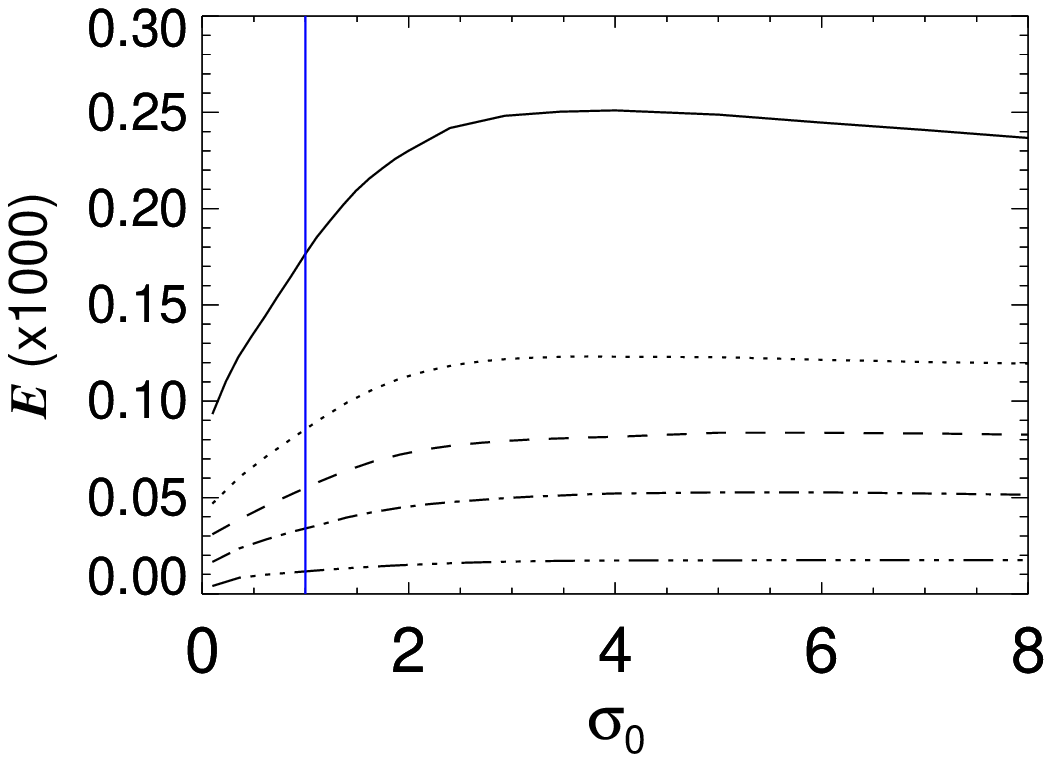}\hspace{-0.5cm}
        }
        \subfigure{
            \includegraphics[width=70mm]{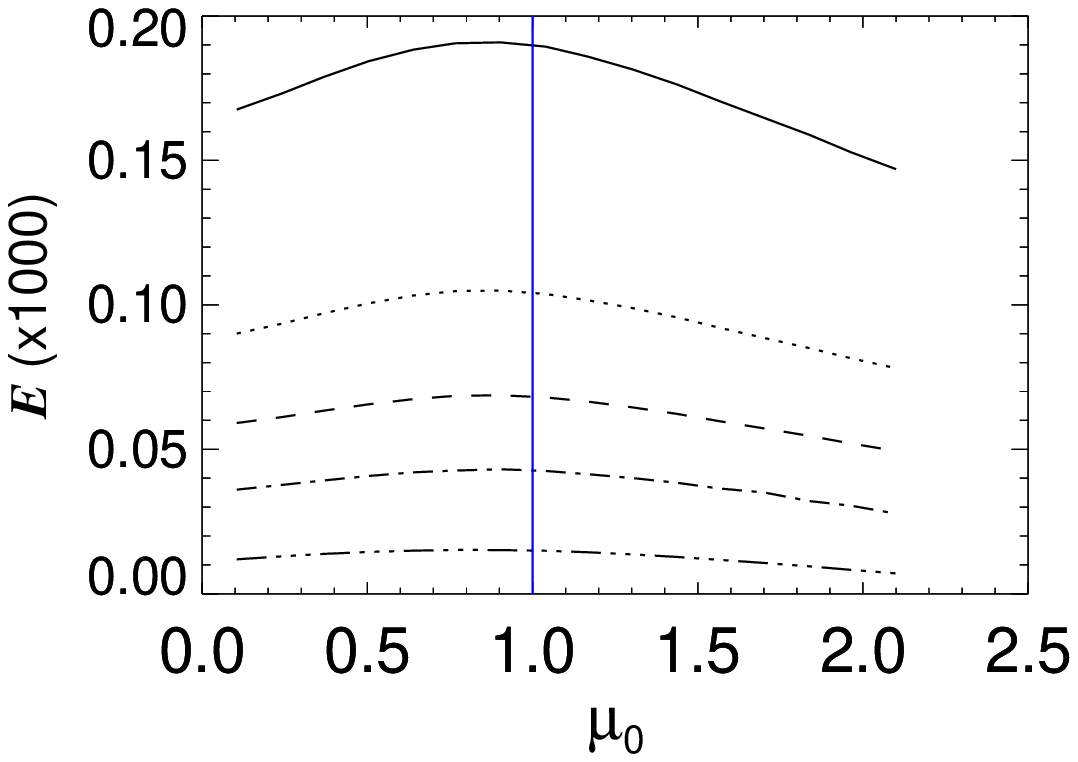}
        }

    \end{center}
    \caption{
        Efficiency plotted against the mixture density parameters. The vertical line in the \textit{bottom left} panel represents the magnitude of the variance of the target distribution, whereas in the \textit{bottom right} it represents the magnitude of the target mean. The different linestyles represent the dimension of the target distribution sampled, with the simplest 2-D distribution being the most efficiently sampled. As expected the 8-D target distribution is most costly to sample.}
   \label{fig:Nmus_Nsigmas}
\end{figure}

The first set of parameters that we optimise are the mixture density parameters, $\mu_0,\mu_1, \sigma_0$ and $\sigma_1$. In Figure \ref{fig:Nmus_Nsigmas} we illustrate how efficiency varies with these parameters in different dimensions ($D= 2, 3, 4, 5, 8$). In all cases the efficiency is greater in lower dimensions because of the smaller number of parameters to be sampled; the scaling with dimension
is studied further in Section \ref{sec:optimNs}.\\

We observe that the efficiency depends very weakly on the parameters $\mu_1$ and $\sigma_1$ in all dimensions, varying by less than 20\% and 40\% respectively over the parameter ranges considered. The slightly preferred value of zero for each of these parameters indicates that larger deviations from the true mean and variance are favoured less. The dependence of efficiency on $\mu_0$ is also relatively weak, with the efficiency peaking at the target value of one as expected, and falling off by less than 20\% over the range considered. This indicates that even though positioning components in high probability regions of the target parameter space is preferred, the penalty for
not achieving this is not significant in the case of a Gaussian target distribution.\\

There is a stronger dependence of efficiency on $\sigma_0,$ with a factor of three difference between efficiency over the 
range of $\sigma_0$ considered. At large values of $\sigma_0,$ relative to the target $\sigma,$ the efficiency is roughly constant indicating that there is only a weak penalty for choosing larger widths of the components. However, the efficiency drops more significantly at smaller values of $\sigma_0$ indicating that it is preferable to always choose a larger width for the initial
covariance of the components. This conclusion is consistent with the MCMC optimisation study in \cite{Dunkley04} but with a weaker dependence of efficiency on $\sigma_{0}$ in our case.

\subsection{Number of components}
\label{sec:optimiseNc}
\begin{figure}
\centering 
\includegraphics[width=150mm]{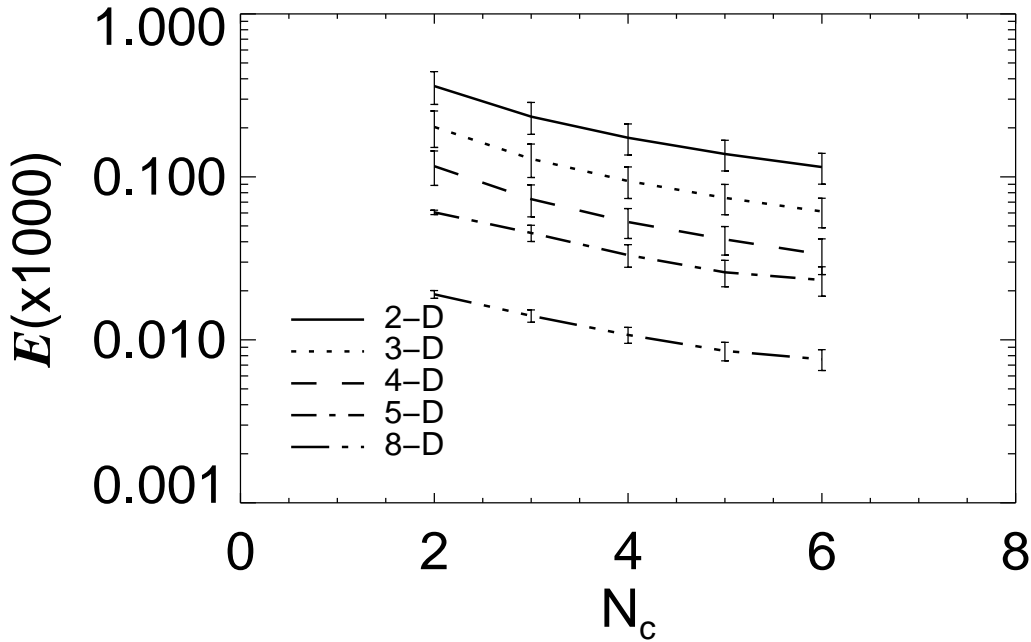}
 \vspace{-0.5cm}
\caption{The efficiency against the minimum number of components, $N_{c}$, needed for convergence. Data points including $1\sigma$ errors are shown along with fitting curves for all dimensions. Errors are obtained from the standard deviations of the thirty realisations.}
\label{fig:D_NeffEff_2D1}
\end{figure}

We now study the relationship between efficiency and number of components, $N_c$. All components are initialised with equal weights.
In Figure \ref{fig:D_NeffEff_2D1} we illustrate how efficiency depends on the number of components in different dimensions. We find that a 
power law fit, $E \propto N_c^{-1},$ is a good description of how efficiency scales with the number of components. This indicates that 
for a multivariate Gaussian target distribution it is always optimal to choose the least number of components i.e., $N_{c}^{\rm opt} = 2,$  because each component is able to adapt to the target distribution. However, this need not be the case for more complex target distributions as we will demonstrate in section \ref{sec:bananastudy}. 

\subsection{Sample size}
\label{sec:optimNs}
\begin{figure}
     \begin{center}
        
        \subfigure{
            \includegraphics[width=85mm]{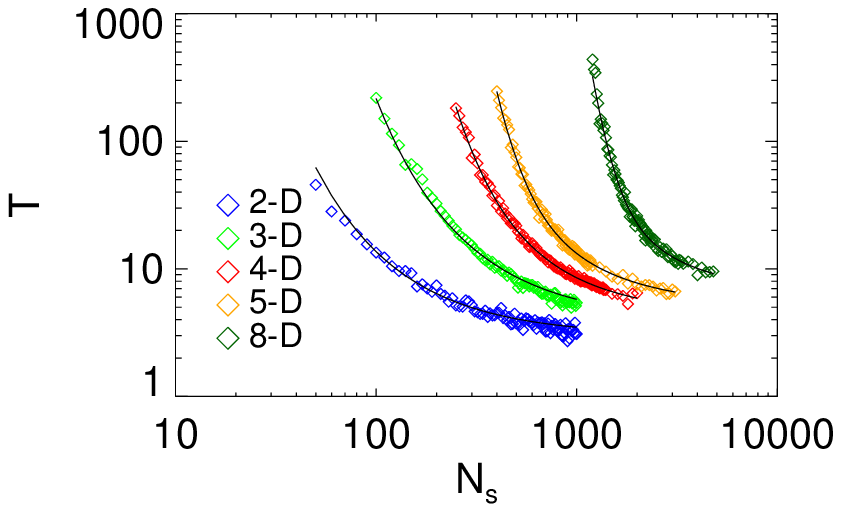}\hspace{-0.5cm}
        }
        \subfigure{
           \includegraphics[width=85mm]{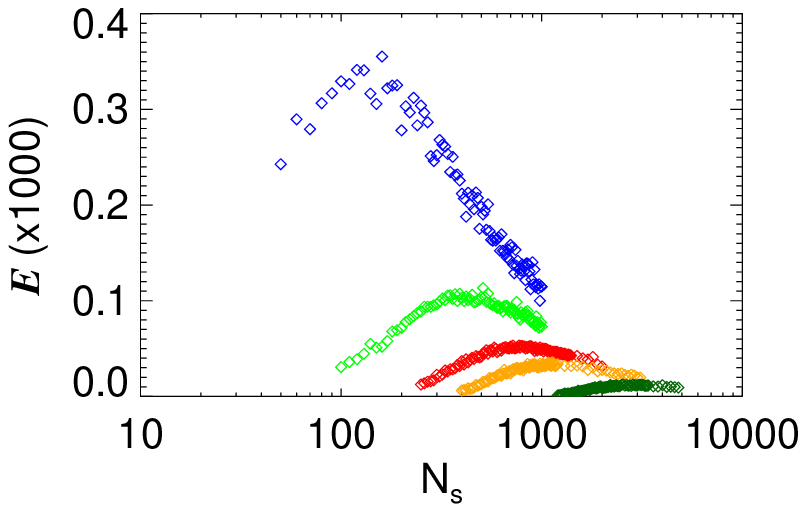}
        }
    \end{center}
    \caption{
        \textit{Left}: The minimum number of iterations required for convergence as a function of sample size, $N_s$, for various dimensions of the target distribution distinguished by the different coloured symbols. Solid lines are fitting curves derived from equation (\ref{eqn:Tfit}). \textit{Right}: The corresponding efficiency against the sample size.}
   \label{fig:NiteffVsN_VaryDim}
\end{figure}

We next consider efficiency as a function of sample size and derive the optimal choice of $N_s$ for a given dimension. The optimal sample size, $N_{s}^{\rm opt}$, should be large enough to yield the required ESS while simultaneously minimising the computational cost. We find that the minimum number of iterations, $T$, required for convergence decreases as a function of sample size, as indicated in the left panel of Figure \ref{fig:NiteffVsN_VaryDim}. To quantify this relationship we used a function of the form 
\begin{equation}
 T(N_s)=a_{1}\exp\left[ \frac{b_1}{N_s} \right] + a_{2}\exp\left[ \frac{b_2}{N_s} \right],
\label{eqn:Tfit}
\end{equation}
where $a_{1},a_{2},b_{1}$ and $b_{2}$ are fitting parameters, to fit the simulated data, with the best-fit shown in the left panel of Figure \ref{fig:NiteffVsN_VaryDim}.\\ 

\begin{figure}
     \begin{center}
        
        \subfigure{
            \includegraphics[width=70mm]{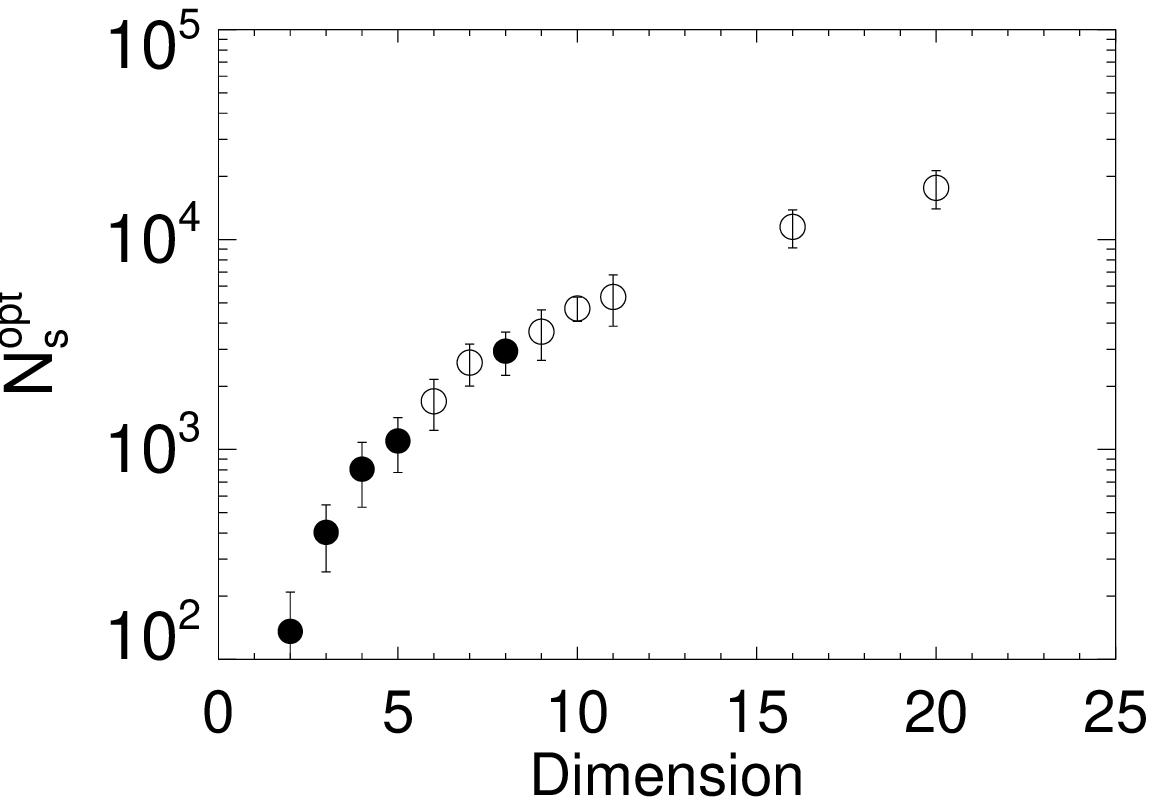}\hspace{-0.5cm}
        }
        \subfigure{
           \includegraphics[width=70mm]{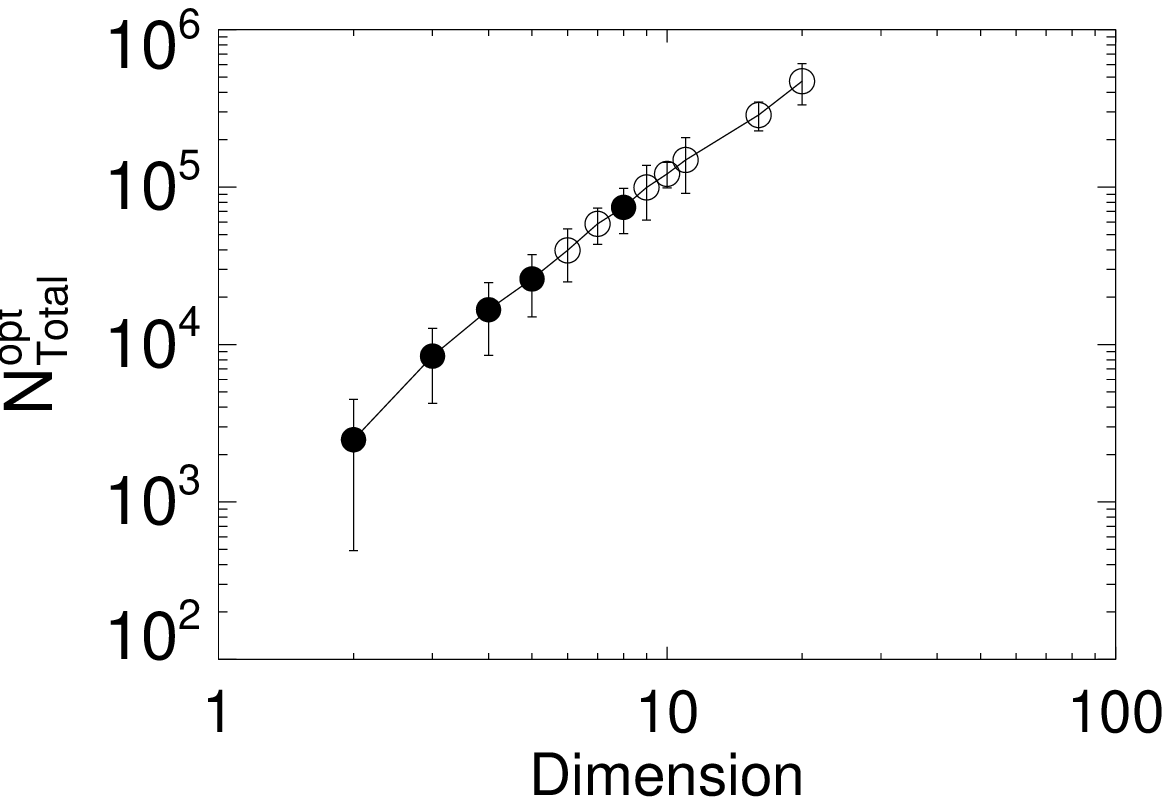}
        }
    \end{center}
    \caption{
        \textit{Left}: The optimal sample size against dimension for multivariate Gaussian distributions, shown with closed circles for the full algorithm parameter simulation. The open circles are for the restricted algorithm parameter simulation. Simulation data is shown with $1\sigma$ errors calculated using thirty realisations. \textit{Right}: The optimal cost of sampling Gaussian distributions as a function of dimension using the optimal sample size from the left panel of this figure, and optimal number of components, $N_{c}^{\rm opt}=2$. The solid line is derived from the fit for $N_s^{opt}$ and the optimal number of iterations and components.}
   \label{fig:Nopt_vs_dim}
\end{figure}

We also find that the number of components remains approximately constant at each iteration, such that the dependence of efficiency on sample size is $E^{-1} \propto N_s\, T(N_s),$ which can be used to find the optimal solution. The dependence of efficiency on sample size is illustrated in the right panel of Figure \ref{fig:NiteffVsN_VaryDim}, in which we have used the optimal number of components, $N_{c}^{\rm opt}=2$. From this figure it is clear that there is an optimal sample size at which the efficiency peaks, and this optimal sample size is greater in higher dimensions. The dependence of optimal sample size, $N_{s}^{\rm opt}$ on dimension, $D,$ is shown in the left panel of Figure \ref{fig:Nopt_vs_dim}, with a simple power law, $N_{s}^{\rm opt} = \gamma D^{\beta}$, providing a good fit to the data. To improve the quality of the fit we added simulated data for additional dimensions, in which we kept $\sigma_1$ and $\mu_1$ fixed to their (weakly) optimal values. This saves on computational time particularly in higher dimensions. The data points for the restricted simulations are indicated as open circles in Figure \ref{fig:Nopt_vs_dim}. We find that the relation is well fit with a power law index, $\gamma \approx 49.1558 \pm 15.53,$ and a normalisation, $\beta \approx 1.964 \pm _{0.03}^{0.049}$. This indicates that there is a quadratic relationship between $N_{s}^{\rm opt}$ and $D$, which can be understood by the number of parameters in the target distribution, corresponding to the number of elements in the covariance matrix, scaling as $D^2$.\\

The above scaling of $N_{s}^{\rm opt}$ with $D$ allows us to determine how the optimal efficiency, $\Eff$, scales with dimension. We found that there is a weak dependence on dimension of the optimal number of iterations, $T^{\rm opt},$ with approximately 10 iterations required for convergence. This means that the computational cost, or inverse efficiency, scales quadratically with dimension, i.e., $E^{-1} = \gamma \, D^\beta \, T^{opt} \,N^{opt}_c \approx 1000 \, D^2,$ as shown in the right panel of Figure \ref{fig:Nopt_vs_dim}. This quadratic scaling is consistent with the computational cost of the MCMC algorithm. Even though the inverse MCMC efficiency defined in \cite{Dunkley04} scales linearly with dimension, this efficiency is defined with respect to an uncorrelated chain. The convergence cost of an uncorrelated chain introduces an additional linear scaling with dimension, which makes the overall inverse efficiency scale quadratically with dimension, as can be seen in Table 2 of \cite{Dunkley04}.

\section{More complex target distributions}
\label{sec:complextarg}
We now study how efficiency degrades for non-ideal target distributions. To this end we investigate whether the mixture densities can adapt to reconstruct more complex distributions. The distributions we consider are the banana shaped and bimodal target distributions.

\subsection{Banana distribution}
\label{sec:bananastudy}
The banana shaped distribution has a strong degeneracy between parameters that are twisted. The target distribution for $D$ random variables, $x_{i}$, for $i=1,\cdots,D$, is a multivariate Gaussian distribution in all dimensions, except $x_D$, which is twisted in the following manner, 
\begin{equation}
 \left( x_{1},\cdots,x_{D-1},y \right) \sim \mbox{N}\left( \mathbf{0},\Sigma \right)
\end{equation}
with the random variable $y$ incorporating the twist such that,
\begin{equation}
 y=x_D + B \left(  \frac{x_{D-1}^2}{\sigma_{1}^2} -1 \right) +\cdots+B\left( \frac{x_{1}^2}{\sigma_{1}^2} -1 \right).
\end{equation}
We have found that for large twists i.e., large values of $B$, simulations in dimensions greater than two are computationally expensive and do not converge in a reasonable time. Hence, we restrict our study of the banana distribution to two dimensions, to investigate how the degree of twist degrades the efficiency. We set $\sigma_{1}^{2}=10$ with covariance matrix $\Sigma=\mbox{diag}(\sigma_{1}^2,\sigma_{1}^2,\cdots,1)$. The parameter $B$ measures the curvature of the banana distribution. In this study, we only consider the parameters $\sigma_{0}$, $\mu_{0}$, $N_{s}$ and $N_{c}$ as $\sigma_{1}$ and $\mu_{1}$ are found to have little effect on efficiency.\\ 

\begin{figure}
\begin{center}
\includegraphics[width=130mm]{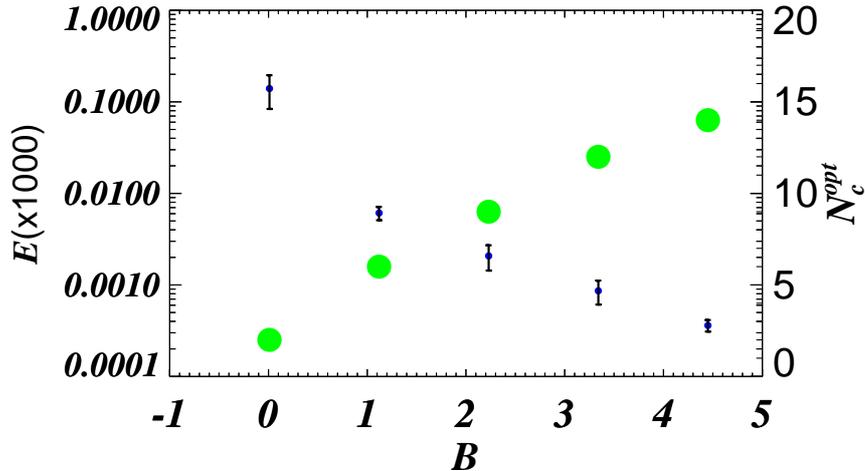}
\end{center}
\caption{The optimum number of components (filled green circles) required to sufficiently sample the target parameter space of a banana shaped distribution as a function of the twist parameter, $B$. The corresponding efficiency is shown by filled blue circles with $1\sigma$ error bars obtained from the standard deviation of thirty realisations. Efficiency is obtained by averaging over all other algorithm parameters varied in the simulation. }
\label{fig:B_Nc2DPhD}
\end{figure}

Components that struggle to sample the target distribution and remain stuck in very low density regions will be discarded. The sampler can be modified such that components that are discarded can be \textit{revived} by positioning them close to the component with highest weight, $\alpha_{j,\textrm{max}}$, and giving them the same covariance matrix as that component. The weights for all components can then be distributed equally. This is needed for more complex distributions and we implement this method in Section \ref{sec:CosmoProblem}. Other reasons that may cause the components to be discarded are due to their covariance no longer satisfying the criteria for being positive-definite, i.e. $\lambda_{i} \leq 0$ for eigenvalue $\lambda$ corresponding to dimension $i$. This is most likely to occur in degenerate regions of the target parameter space.\\

The optimum number of components, $N_{c}^{opt}$, required for convergence increases with $B$ as shown in Figure \ref{fig:B_Nc2DPhD}. This is expected since the target parameter space becomes more degenerate with larger tails of low density. More components are therefore required to sufficiently sample these regions. The target distribution is only successfully sampled for $B < 5$; for larger values of $B$ the distribution tails become too narrow to sample from adequately. The maximum efficiency obtained for each corresponding $N_{c}^{opt}$ is shown in Figure \ref{fig:B_Nc2DPhD}, which illustrates how the performance is degraded for increased values of $B$.

\subsection{Bimodal distribution}
\label{sec:bimodalstudy}
Distributions with more than one peak can pose a problem for PMC, specifically those with narrow peaks connected by regions of very low probability. We consider the symmetric bimodal distribution
\begin{eqnarray}
 p(\mathbf{x}) 	&\propto& \exp\left[ -\frac{1}{2}(\mathbf{x}-\boldsymbol{a})\Sigma^{-1} (\mathbf{x}-\boldsymbol{a})^T \right] \nonumber \\
	&+& \exp\left[ -\frac{1}{2}(\mathbf{x}+\boldsymbol{a})\Sigma^{-1} (\mathbf{x}+\boldsymbol{a})^T \right],
\end{eqnarray}
with modes situated at $\pm \boldsymbol{a}$ where $\boldsymbol{a}=(a_1,a_2,\cdots,a_D)$ in $D$ dimensions. We set the covariance matrix to be simply the identity matrix, $\Sigma = I$. For simplicity we fix the peak separation to be the same in all dimensions, with $a_i=a$, and only study the two dimensional case.\\
 
\begin{figure}
\begin{center}
 \includegraphics[width=130mm]{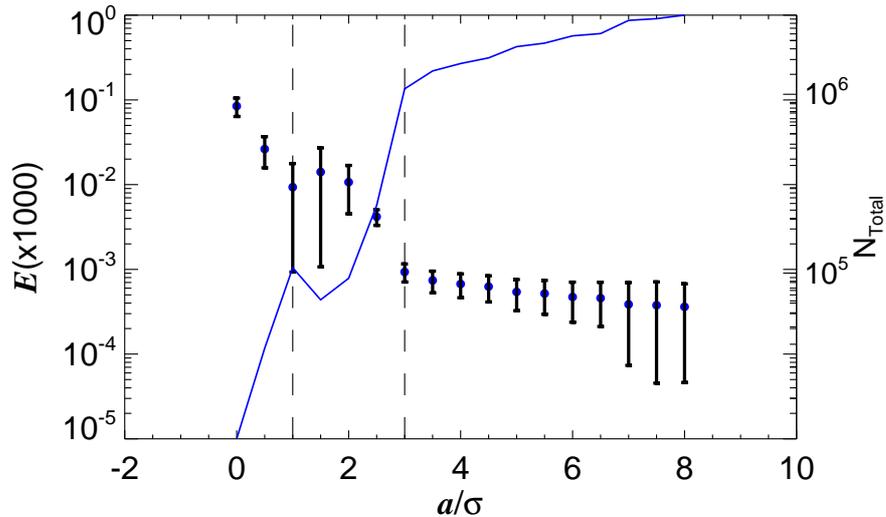}
\end{center}
\caption{Plot of efficiency (filled circles) for the bimodal distribution against the separation of the modes, $a$, scaled by the width of the peaks. The corresponding cost, $N_{Total}$, of the simulation is shown by the solid line. The region $1 \leq a \leq 3$ is separated by vertical dashed lines. }
\label{fig:a_eff_Ntotal}
\end{figure} 

The relation between $\Eff$ and $a$ is shown in Figure \ref{fig:a_eff_Ntotal}. For small enough $a$ the target distribution closely resembles a Gaussian distribution and the results from Section \ref{sec:OptimisePMC} are relevant. However, for larger $a$ the components tend to sample both modes until each component adapts to one mode. This increases the number of iterations required for convergence. For large enough $a$, the modes become distinct and are separated by a region of low probability. In this case the PMC algorithm exhibits the lowest efficiency. Additionally we are faced with the problem of false convergence, in the case that components adapt to only one peak and produce a satisfactory ESS when in fact the algorithm has not sampled the second peak. This type of distribution presents a challenge to the PMC algorithm and the convergence criteria must be appropriately modified in order to overcome this challenge.

\section{Application to cosmology}
\label{sec:CosmoProblem}
We now apply the optimised PMC algorithm to a cosmological problem, namely, constraining the contribution of isocurvature perturbations in the early universe in addition to the usual adiabatic perturbations that are assumed to have seeded structure formation. The cosmological application allows us to optimise the performance of the PMC sampler in a more realistic parameter estimation problem. We use cosmic microwave background (CMB) data from the Wilkinson Microwave Anisotropy Probe (WMAP) 9 year release \citep{Hinshaw2013} to constrain a combination of adiabatic and isocurvature mode amplitude parameters together with a set of six cosmological parameters. We consider a general admixture model with an admixture of the adiabatic mode correlated with all three isocurvature modes that adds nine mode amplitude parameters. The constraints presented here are the most up to date in the literature from WMAP 9 year data for the general admixture model. In a subsequent study we plan to extend the application of these methods to the latest temperature and polarization data from Planck. 

\subsection{CMB likelihood, models and parameters}
\label{sec:data_mod_like}

We calculate the likelihood of the WMAP 9 year data for a given model using the publicly available Version 5 WMAP 9 code \citep{Bennet2013, Larson2011, Dunkley2009, Hinshaw2003, Verde2003}. The code uses the CMB temperature and polarisation angular power spectrum and covariance from the WMAP 9 year release \citep{Hinshaw2013}. The likelihood for the temperature spectrum is split into two parts, namely a low multipole range ($2 \leq \ell \leq 32$) and a high multipole range ($33 \leq \ell \leq 2000$). Similarly the likelihood for the polarisation spectrum for EE, TE and BB modes is split into $2 \leq \ell \leq 24$ and $25 \leq \ell \leq 2000$ ranges. We use the publicly available package CAMB \citep{Lewis2000} to compute the CMB spectra for a given model, which is then fed into the likelihood code. The model parameters we consider here are the standard six cosmological parameters for the flat $\Lambda\mbox{CDM}$ model, namely the baryon density $(\Omega_{b}h^2)$, the cold dark matter density $(\Omega_{c}h^2)$, the dark energy density $(\Omega_{\Lambda})$, the optical depth to re-ionization $(\tau )$, the scalar spectral index $(n_s )$, and the amplitude of the primordial power spectrum $(A_s )$. We set flat uniform priors over a large enough range of interest for the cosmological parameters, apart from $A_s$, which has a positivity prior. The allowed parameter ranges are shown in Table \ref{table:cosmoparampriors}.\\
 
\begin{table}
\centering
\scalebox{0.8}{
\begin{tabular}{|c|c|c|c|}
\hline\hline
Symbol		&	Description	& Minimum	& Maximum	 \\
\hline
$\Omega_{b}h^2$	& Baryon density	& 0.01	& 0.04	 \\
\hline
$\Omega_{c}h^2$	& Cold dark matter density	& 0	& 0.4	 \\
\hline
$\Omega_{\Lambda}$	& Dark energy density	& 0.5	& 0.9	 \\
\hline
$\tau$		& Optical depth to re-ionization	& 0.01	& 0.2	\\
\hline
$n_s$		& Scalar spectral index	& 0.8	& 1.3	\\
\hline
$A_s$		& Amplitude of the 		& 0	& none	\\
		& primordial power spectrum	&	&	\\
\hline
\end{tabular}
}
\caption{Priors on cosmological parameters in both the adiabatic and admixture model.}    
\label{table:cosmoparampriors}
\end{table}

In the admixture study we consider an adiabatic mode (AD), and four isocurvature modes to model the initial conditions. The four isocurvature modes are: the CDM (CI), baryon (BI), neutrino density (NID), and neutrino velocity (NIV) isocurvature modes. These four isocurvature modes correspond to perturbations in entropy with no perturbation in curvature, and density ratios of different species that are not spatially constant in their perturbations initially. The baryon and cold dark matter isocurvature modes produce the same CMB spectra, hence we only consider the CI mode here, which leaves us with $N_{\rm mode}$ = 4 modes viz. AD, CI, NID, NIV. Previous work in constraining admixture models with adiabatic and isocurvature modes has been presented in \cite{Savelainen2013PhRvD_88f3010S} and \cite{Moodley2004PhRvD_70j3520M}, which comprise a subset of the papers on this topic using the MCMC sampling algorithm. We closely follow the parametrization and methodology of their work in implementing the optimised PMC algorithm.\\

The adiabatic and isocurvature auto- and cross-correlation spectra are shown in \cite{Kasanda_Moodley2014}. The overall CMB anisotropy, $C_{\ell}$, is a sum over normalised mode spectra, $C_{\ell}^{IJ}$, with amplitudes $A_{IJ}$ such that
\begin{equation}
C_{\ell}=\sum_{IJ}A_{IJ}C_{\ell}^{IJ},
\end{equation} 
where the indices $I,J=1,2,3,4$ label the modes AD, CI, NID, NIV respectively. A symmetric dimensionless matrix, with components $Z_{IJ} \propto A_{IJ},$ contains all information on the fractional contributions (in the sense that $\sum_{IJ} Z_{IJ}^2 = 1$) of the auto- and cross-correlation spectra to the overall CMB spectrum. We refer the reader to \cite{Moodley2004PhRvD_70j3520M} for more information on the sampling of the $Z_{IJ}$ matrix elements and the proportionality factor in the previous relation. We also use their definition of the non-adiabatic fraction,
\begin{equation}
f_{\text{ISO}}=\frac{Z_{\text{ISO}}}{Z_{\text{ISO}}+Z_{\left<\text{AD,AD}\right>}}
\end{equation}
with $Z_{\text{ISO}}$ representing the fraction of isocurvature contribution to the data given by
\begin{equation}
Z_{\text{ISO}}=\sqrt{1-Z^{2}_{\left< \text{AD,AD} \right>} }
\end{equation}
for both auto- and cross-correlations.

\subsection{Performance of PMC for adiabatic and isocurvature models}

In this section we study the efficiency of the PMC algorithm in the multi-dimensional parameter space of these admixture models. We focus on how the algorithm performs for a target distribution that results from realistic cosmological data compared to the case of the idealised Gaussian target distribution studied in a previous section.\\

We consider an adiabatic model and an admixture model with the adiabatic and three isocurvature modes. These have respective dimensions of six and fifteen. The adiabatic model parameters are listed in Table \ref{table:cosmoparampriors}, whereas the admixture model contains the first five cosmological parameters listed in Table \ref{table:cosmoparampriors} and ${1\over 2} N_{\rm mode} (N_{\rm mode}+1)$ mode amplitude parameters. Based on our results from the previous section we choose the optimal sample size per iteration, based on these dimensions, to be 2 000 and 10 000 respectively.\\

The choice of the optimal number of components, $N_c,$ is more complicated. For the pure adiabatic model we found that 2 components, the optimal number obtained
in the Gaussian case, suffices. However, when isocurvature modes are included the parameter space becomes more complex with 
banana-like degeneracies introduced. A balance has to be reached between choosing a large enough number of components to allow convergence but not too large a number such that the efficiency is reduced. By systematically increasing the number of components until convergence was reached we found that 10 components worked for the isocurvature mode case. This number of components is consistent with that adopted in \cite{Wraith09}.\\

We found that the more complex isocurvature parameter space necessitated the use of the {\it revive} method discussed in Section \ref{sec:bananastudy} to recover components with negligible weights. Furthermore the complex higher dimensional space of the admixture model causes the conventional ESS to stabilise at very low values because there is significant degeneracy of the importance weights, $\{w_{n}^t \}$ at iteration $t$. To overcome the degeneracy problem of the importance weights, we adopted the Non-linear Population Monte Carlo algorithm presented in \cite{Koblents2012arXiv1208_5600K}. We specifically use the soft clipping transformation introduced in this paper.\\

To initialise the mixture density components we started with a fiducial model with pure adiabatic perturbations that provided a reasonable fit to the data. For the adiabatic case we computed the Fisher matrix about this fiducial model, using an analytic noise model that was well matched to the WMAP 9-year data \citep{Kasanda_Moodley2014}. A scaled inverse Fisher matrix was used as the initial covariance matrix for each of the components with the scaling factor randomly chosen in the range $[1,3]$, following our result for its optimal value in Section \ref{sec:mixturedensities}. For the admixture model we extended the Fisher matrix computed in the adiabatic case to include the isocurvature amplitudes with zero cross-correlations and small auto-correlations ($\sim 10^{-3}$). This choice allowed the mixture density components to be initialised close to the fiducial model, with further PMC iterations allowing the components to adapt to the degenerate directions in the admixture model parameter space.\\

The performance of the PMC algorithm in the adiabatic and admixture parameter space is presented in Table \ref{table:NtotalANDEffISO}, where we tabulate the cost of sampling the underlying distributions compared to the optimal cost for a Gaussian distribution found in Section \ref{sec:OptimisePMC}. We observe that the adiabatic and admixture model runs required approximately three and four times more iterations, respectively, than the optimal number of iterations ($\sim$10) for a Gaussian distribution found in Section \ref{sec:optimNs}, likely due to non-Gaussianity in the target distributions. For the admixture model the cost was increased by an additional factor of 5 arising from the larger number of components used. As discussed above a larger number of components is needed to guarantee convergence in a complex parameter space with banana-like degeneracies.

\begin{table}
\centering
\begin{tabular}{|c|c|c|c|c|c|}
\hline\hline
Model	& D & $N_{it}$	& $N_{total} \,(10^5)$ & $N_{total}^{ideal}\,(10^5)$	& $E$ $(10^{-6})$	 \\
\hline
AD 	& 6 & 26	& 1.09  & 0.36	& 9.16 	 \\ 
\hline 
AD+CI+NIV+NID	& 15	& 41		& 41.0	& 2.25 & 0.24 \\
\hline
\end{tabular}
\caption{The cost and efficiency of sampling a pure adiabatic model and a general admixture model.}   
\label{table:NtotalANDEffISO}
\end{table}

\subsection{Adiabatic model}
\label{ssec:OptPMCcosmoparamestAD}
\begin{figure}
    \begin{center}
       \subfigure{
           \includegraphics[width=60mm]{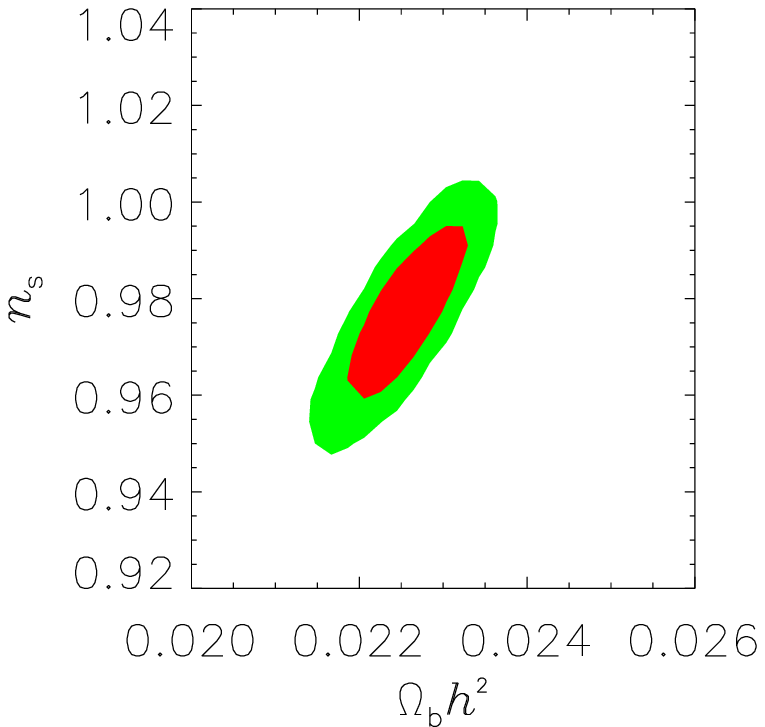}
       }\hspace{-0.15cm}%
       \subfigure{
          \includegraphics[width=60mm]{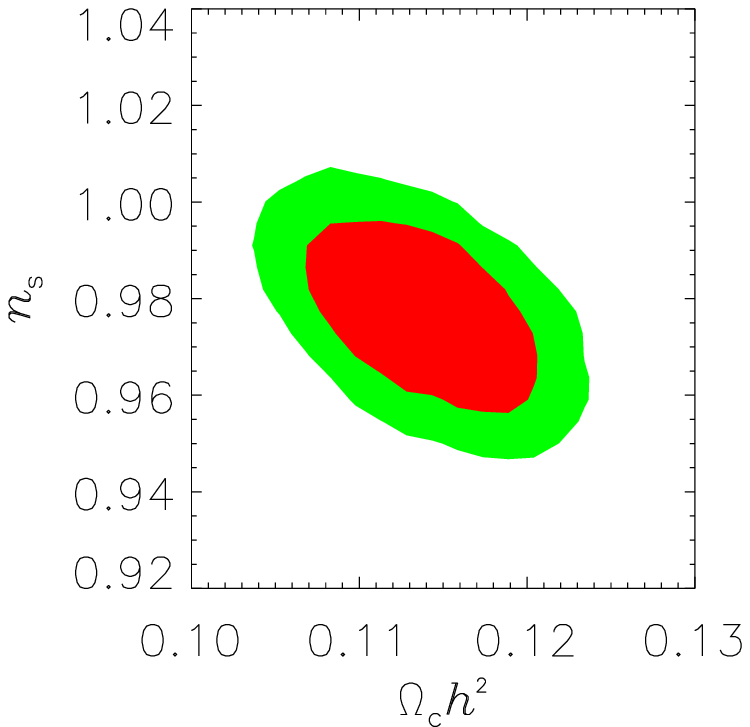}
       }\\ 
       \vspace{-0.5cm}
       \subfigure{
           \includegraphics[width=60mm]{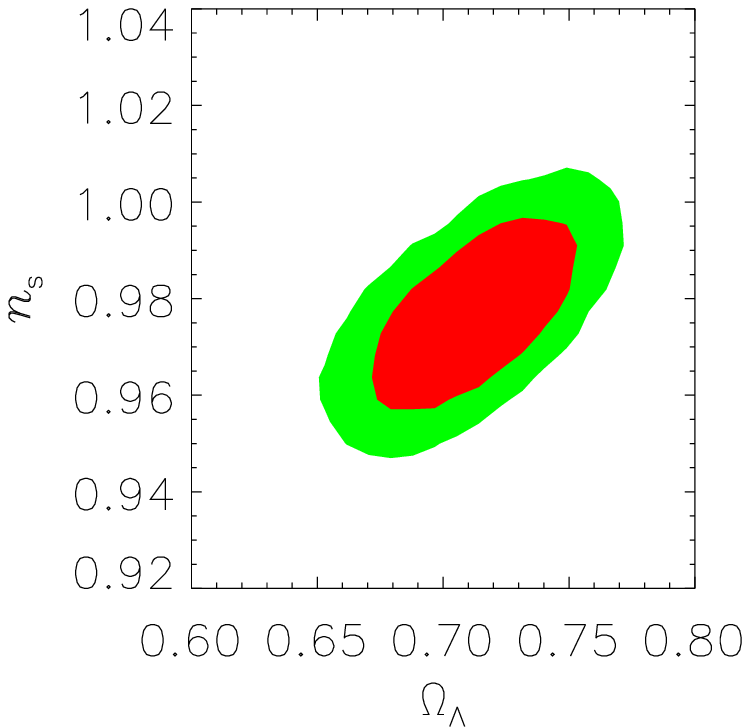}
       }\hspace{-0.15cm}%
       \subfigure{
           \includegraphics[width=60mm]{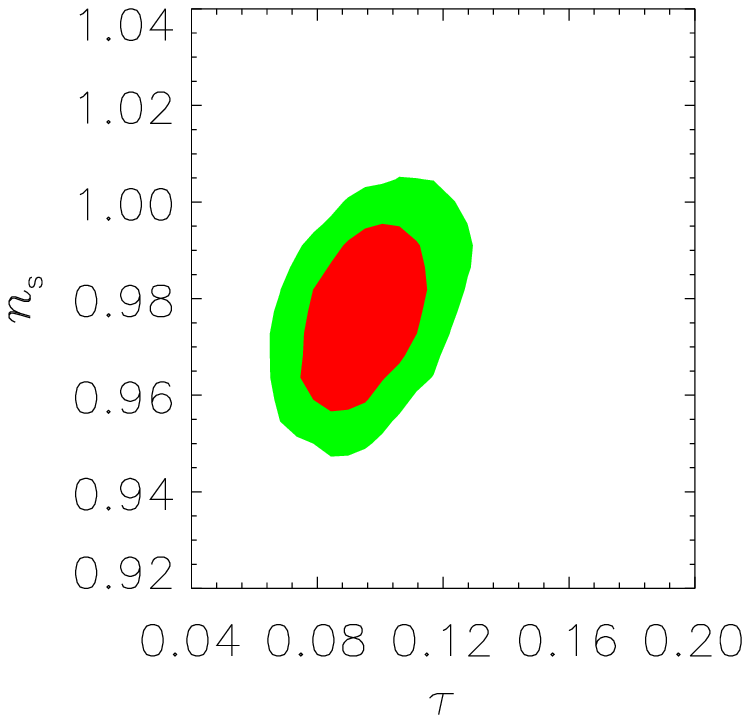}
       }
   \end{center}
   \caption{{\bf Adiabatic model}. Joint distributions between certain parameters are plotted to illustrate the mild degeneracies present in the adiabatic model. Confidence regions of $1\sigma$ and $2\sigma$ contours are shown in the green (lighter) and red (darker) colours respectively.}%
  \label{fig:contourcosmoparams}
\end{figure}

We now study the constraints on a pure adiabatic cosmological model. This case provides a useful test of our PMC code against the results obtained in the literature. In Table \ref{table:EstMLE} we show the marginalised values and 1-$\sigma$ errors on cosmological parameters obtained from the PMC algorithm in comparison to the results from the WMAP team \citep{Hinshaw2013}. There is good agreement between both sets of parameter values, both in the central values and the size of the error bars. In Figure \ref{fig:contourcosmoparams} we show joint distributions between a few selected parameters that illustrate the mild degeneracies present in certain parameter combinations.

\begin{table}
\vspace{0.5cm}
\centering
\begin{tabular}{|c|c|c|}
\hline\hline
Parameter		& PMC analysis 	& WMAP team analysis  \\
\hline
$\Omega_{b}h^2$		& $0.0226 \pm 0.0005$	& $0.0226 \pm 0.0005 $	 \\
\hline
$\Omega_{c}h^2$		& $0.114 \pm 0.005$		& $0.114 \pm 0.005$	 \\
\hline
$\Omega_{\Lambda}$	& $0.71 \pm 0.03$		& $0.72 \pm 0.03 $	 \\
\hline
$\tau$			& $0.10 \pm 0.01$		& $0.09 \pm 0.01$	\\
\hline
$n_s$			& $0.98 \pm 0.01$		& $0.97 \pm 0.01$	\\
\hline
\end{tabular}
\caption{{\bf Adiabatic model}. Central parameter values and 1-$\sigma$ errors obtained using the PMC algorithm compared to a previous cosmological parameter analysis \citep{Hinshaw2013} for WMAP nine-year data.}    
\label{table:EstMLE}
\end{table}

\subsection{Adiabatic mode correlated with all three isocurvature modes}
\label{ssec:Ad_allmode}
    
\begin{figure}
\centering
\includegraphics[width=40mm]{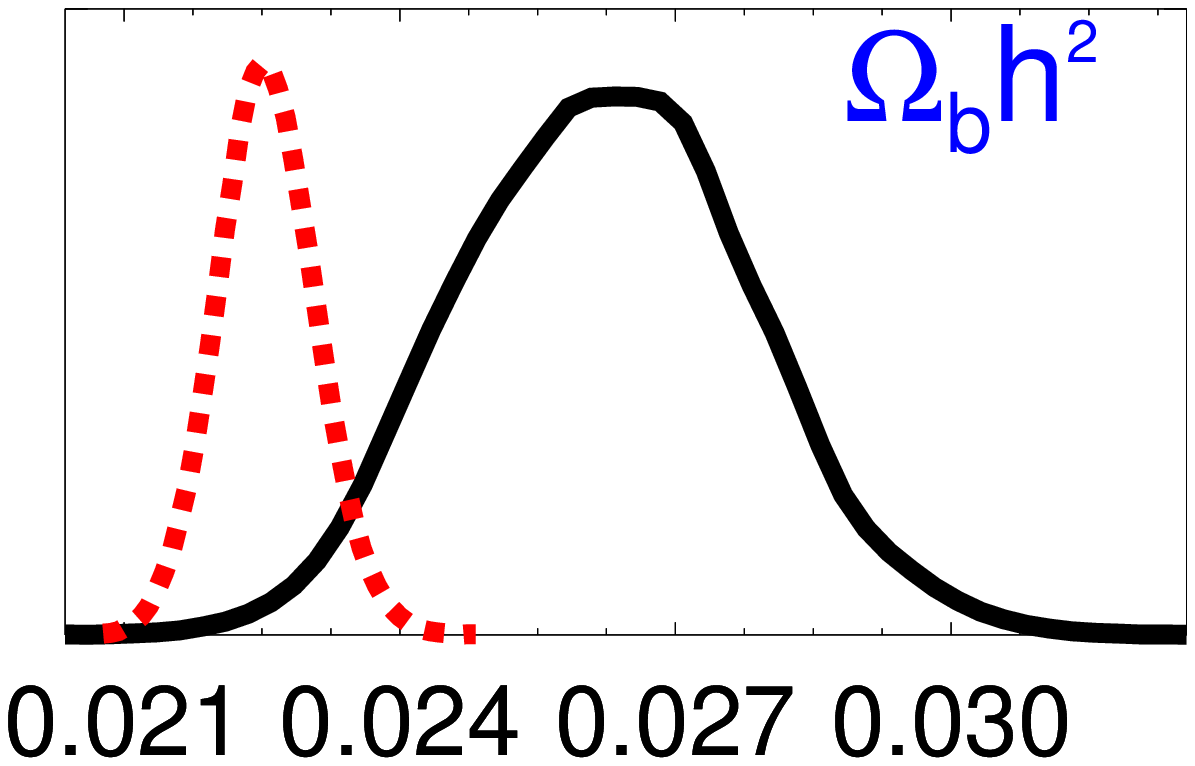}\hspace{-0.75cm}
\includegraphics[width=40mm]{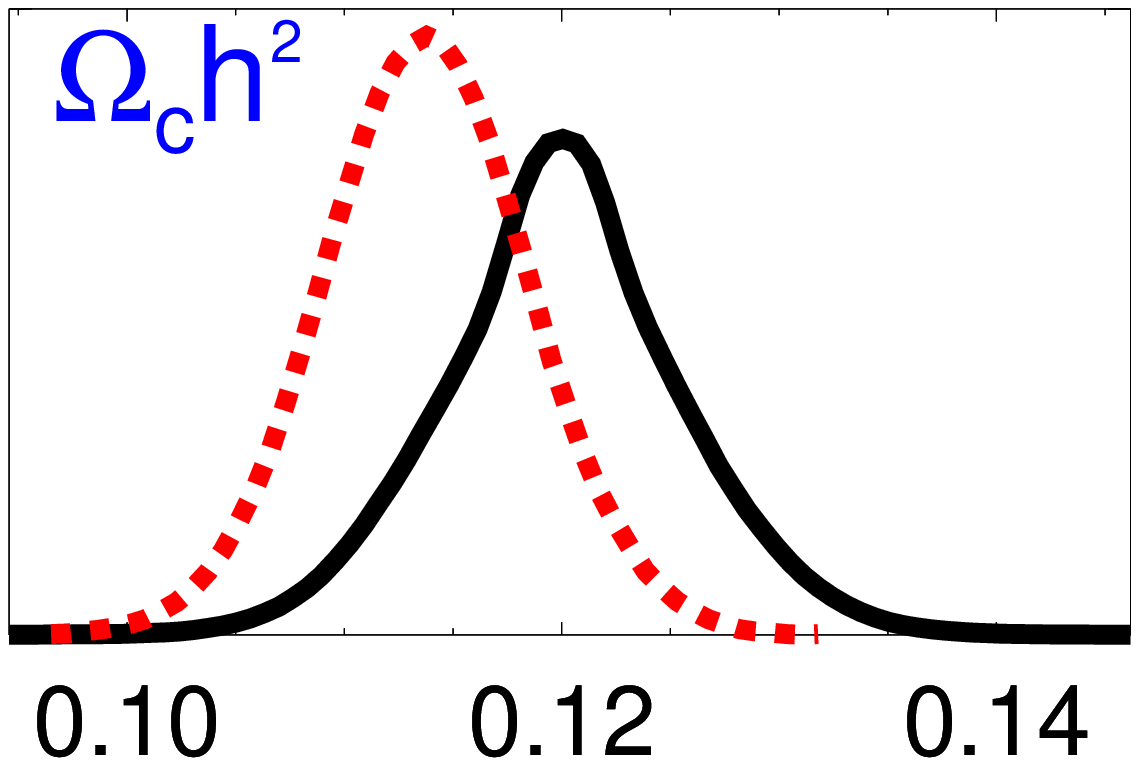}\hspace{-0.75cm}
\includegraphics[width=40mm]{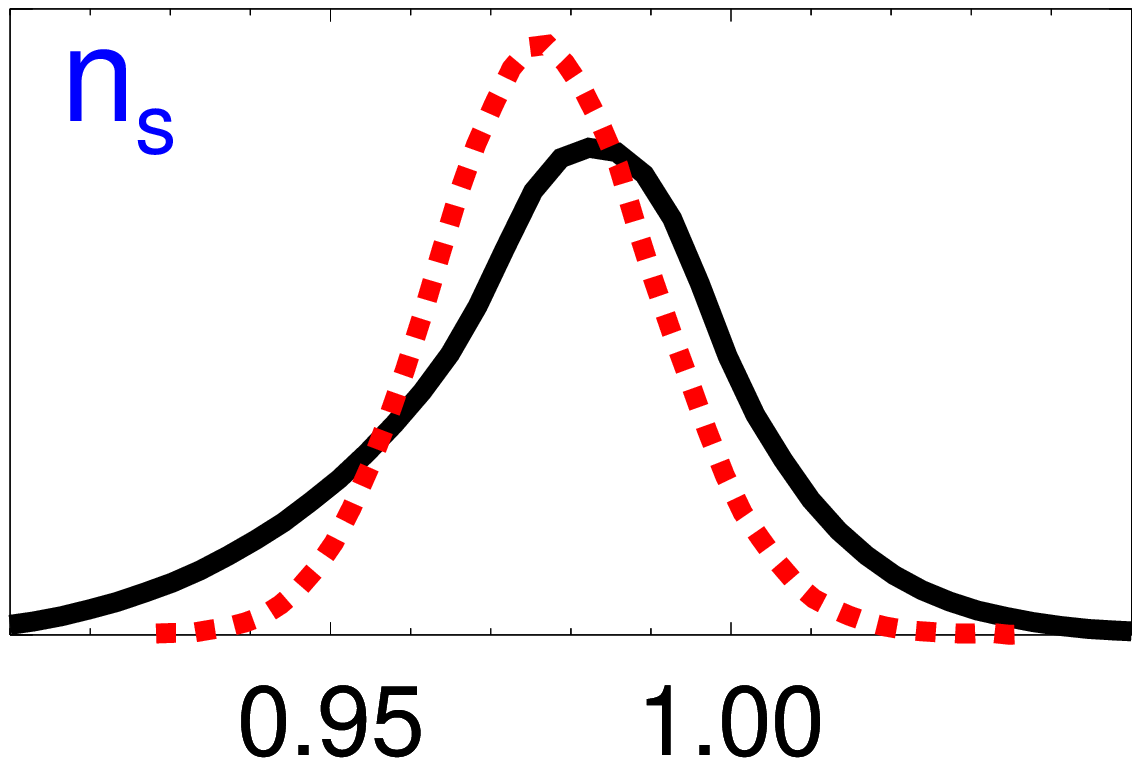}\hspace{-0.75cm}
\includegraphics[width=40mm]{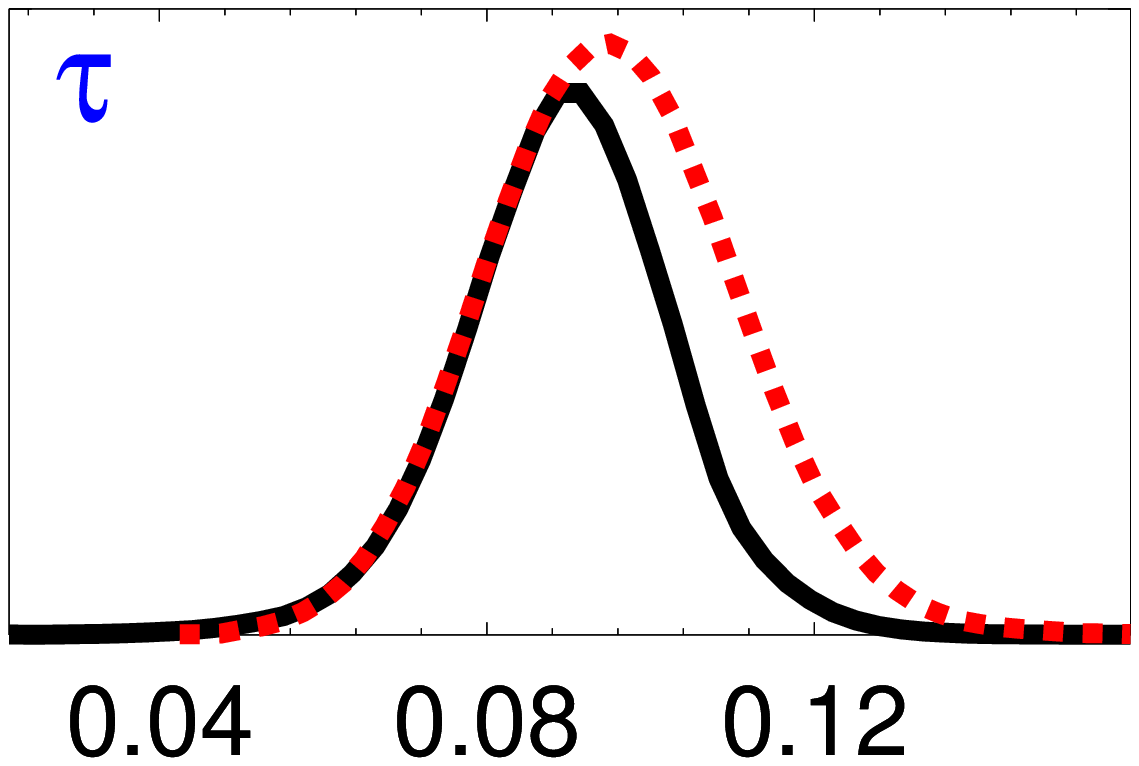}\\
\vspace{-0.1cm}
\includegraphics[width=40mm]{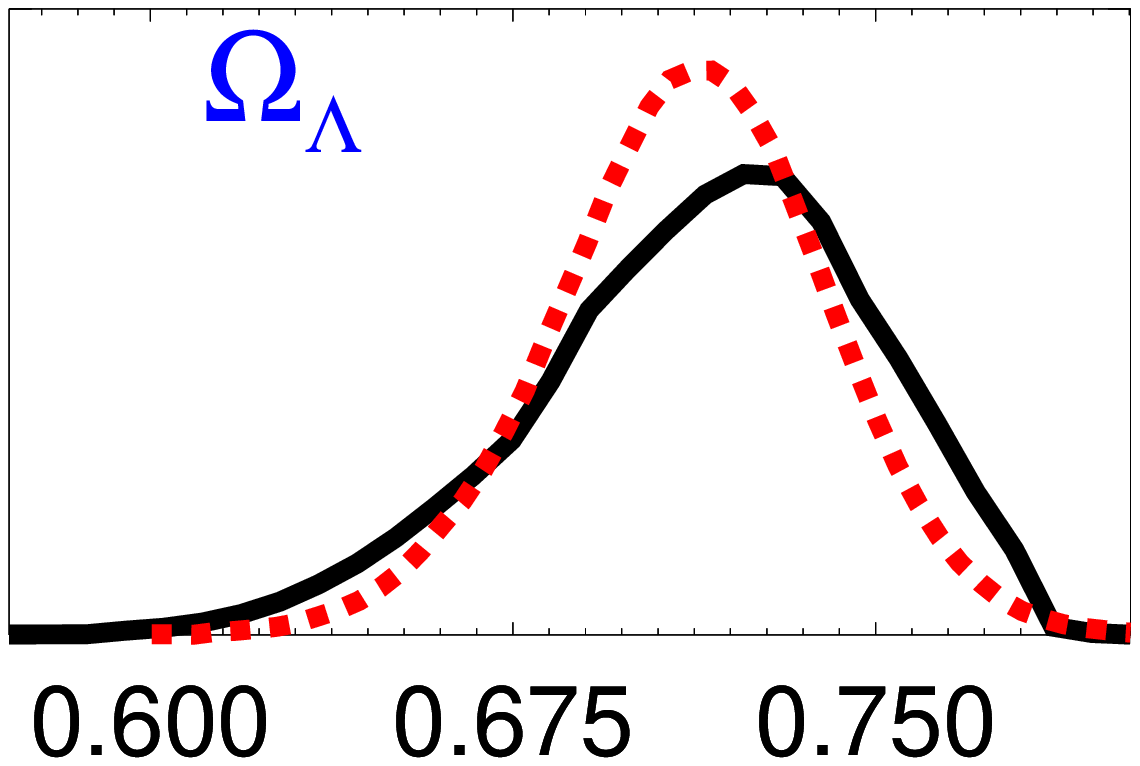}\hspace{-0.75cm}
\includegraphics[width=40mm]{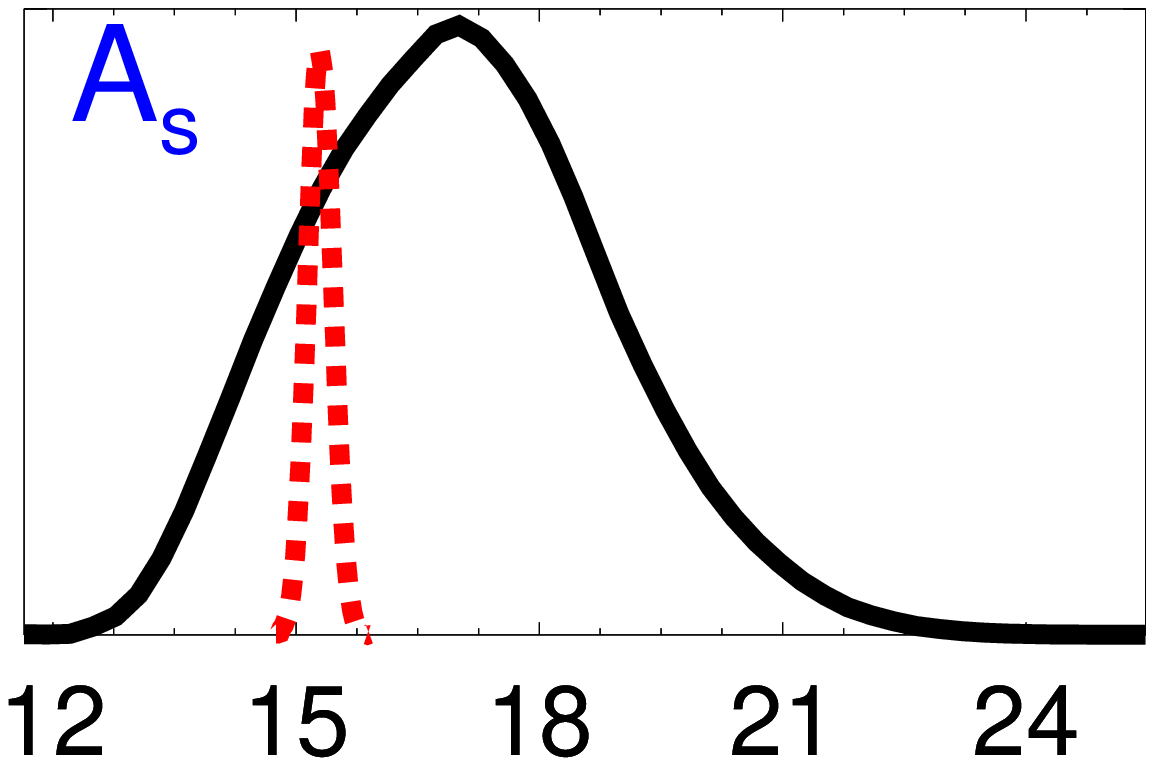}\hspace{-0.75cm}
\includegraphics[width=40mm]{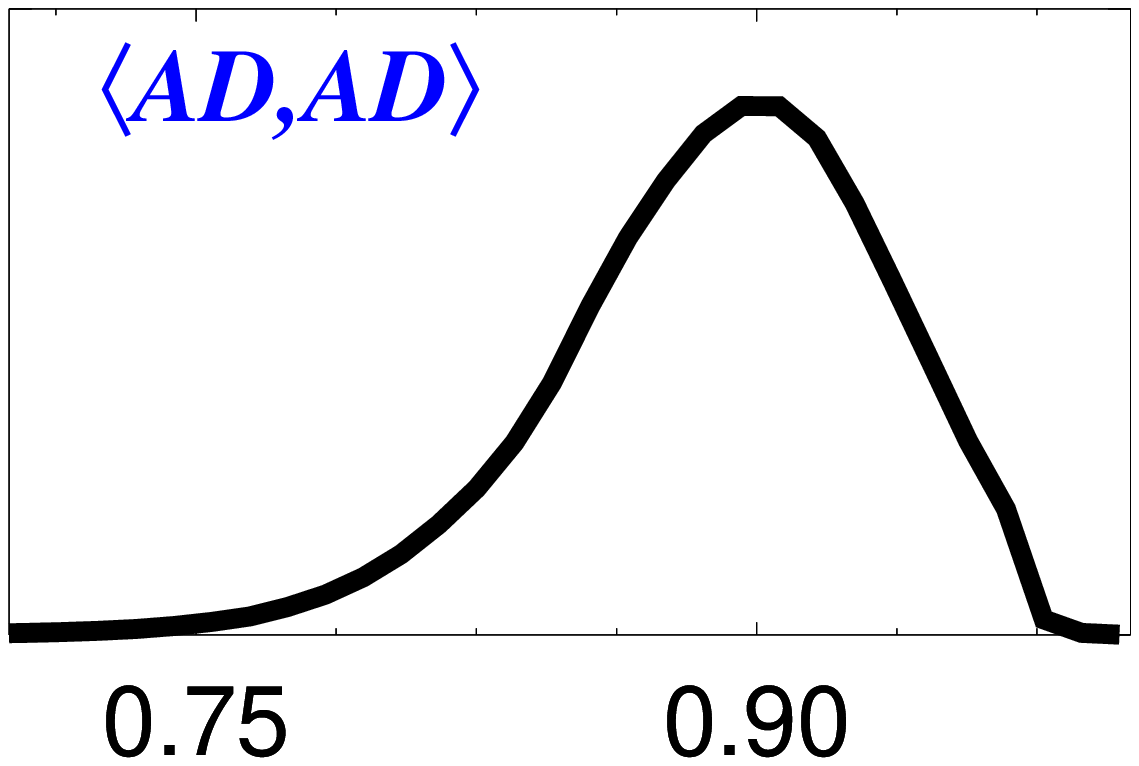}\hspace{-0.75cm}
\includegraphics[width=40mm]{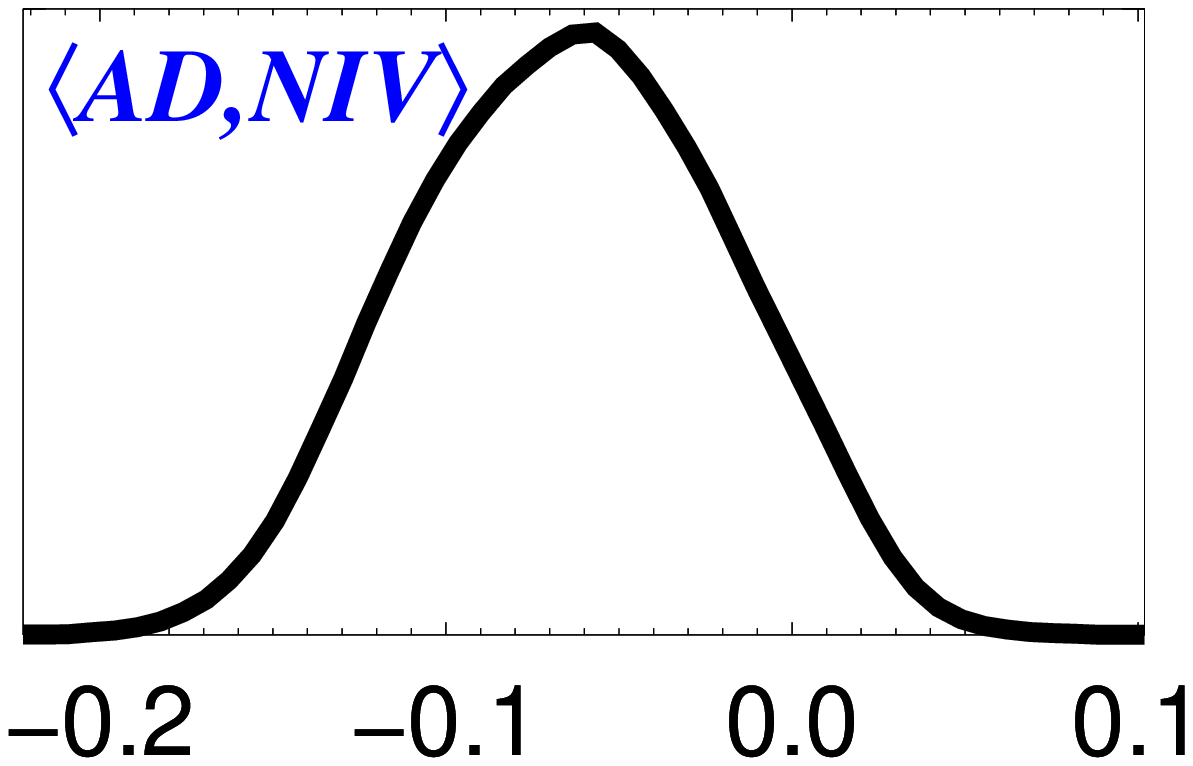}\\
\vspace{-0.1cm}
\includegraphics[width=40mm]{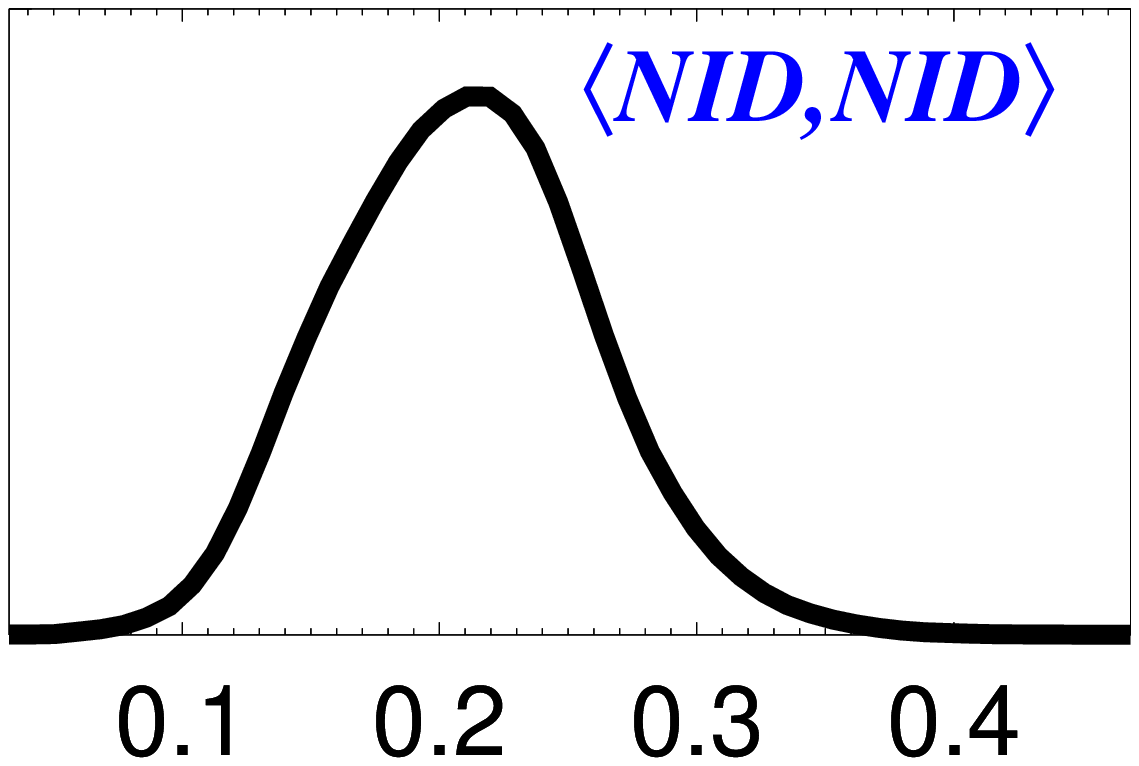}\hspace{-0.75cm}
\includegraphics[width=40mm]{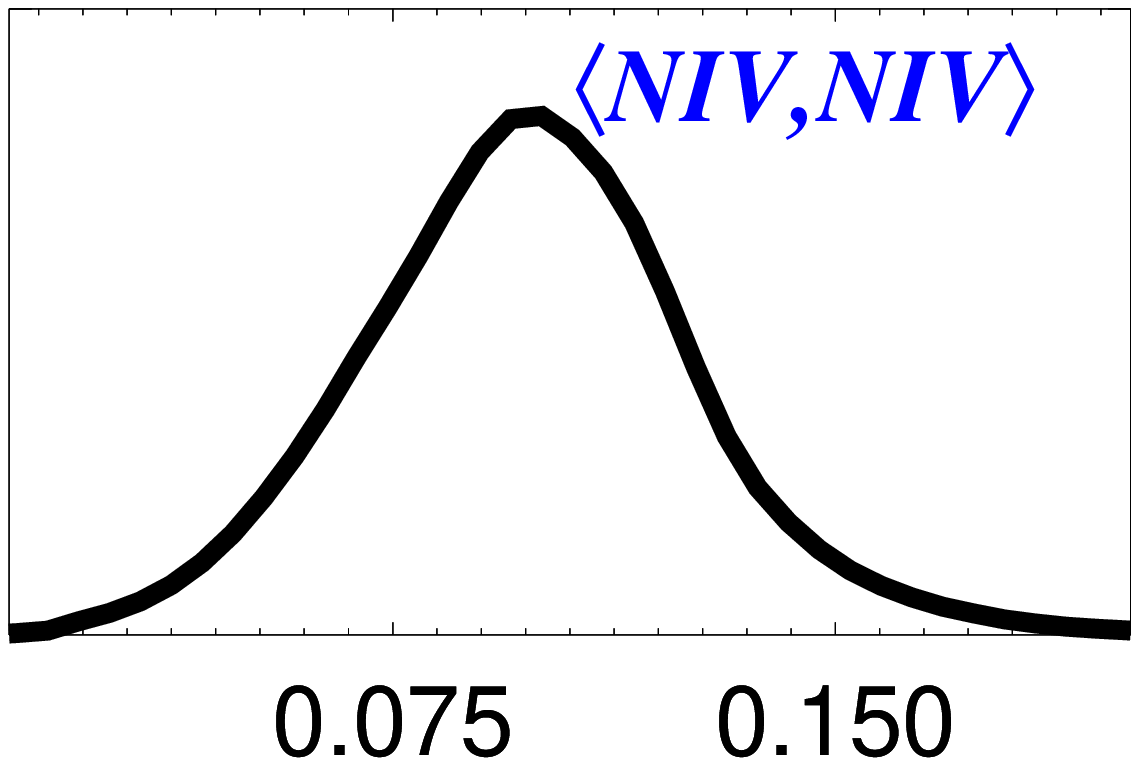}\hspace{-0.75cm}
\includegraphics[width=40mm]{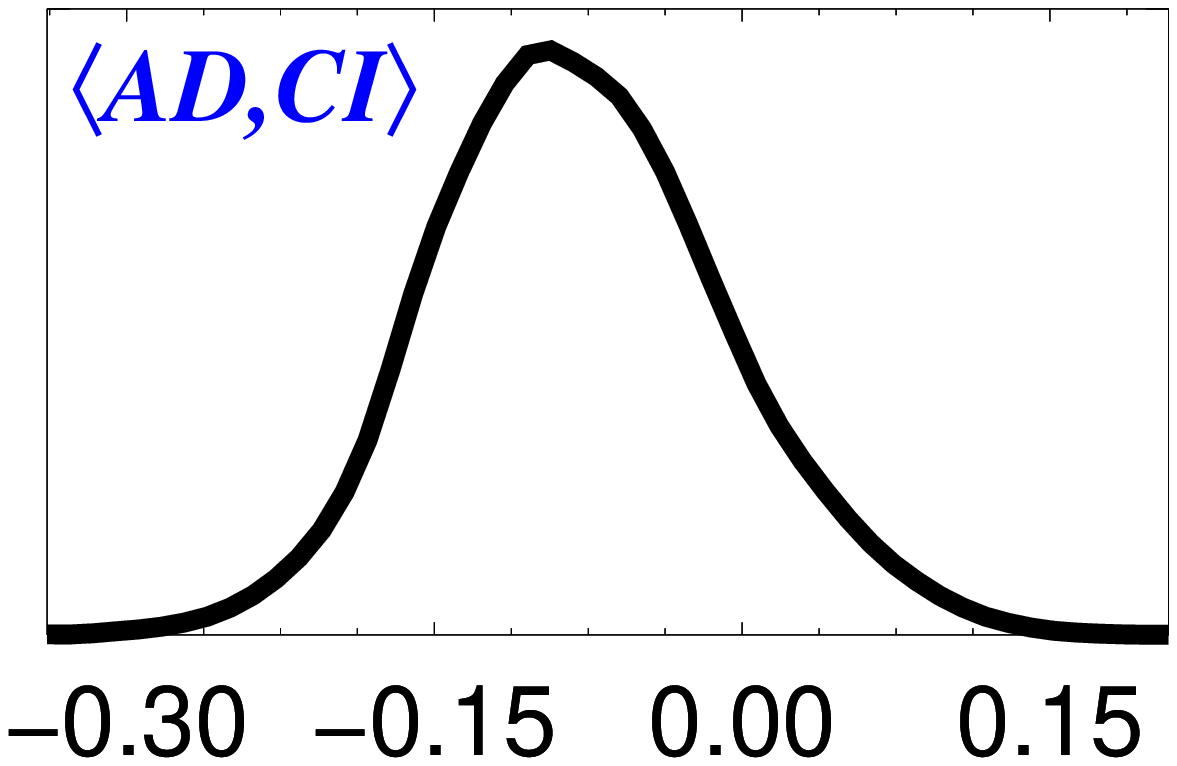}\hspace{-0.75cm}
\includegraphics[width=40mm]{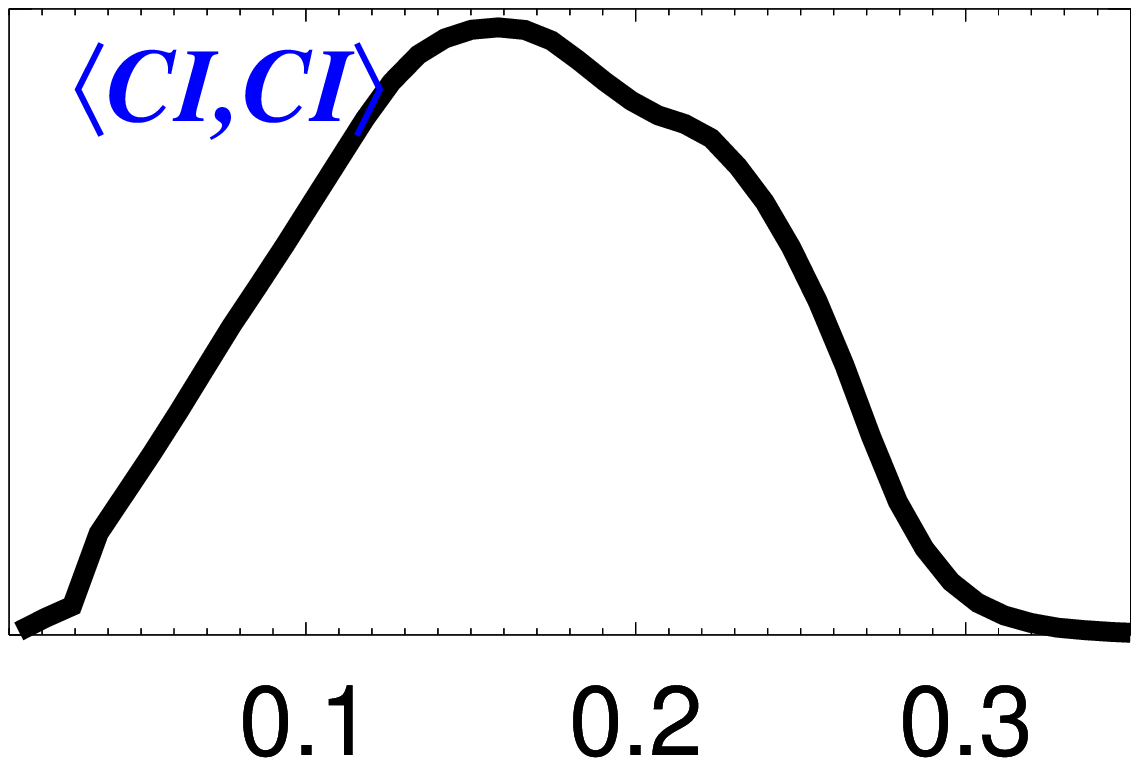}\\
\vspace{-0.1cm}
\includegraphics[width=40mm]{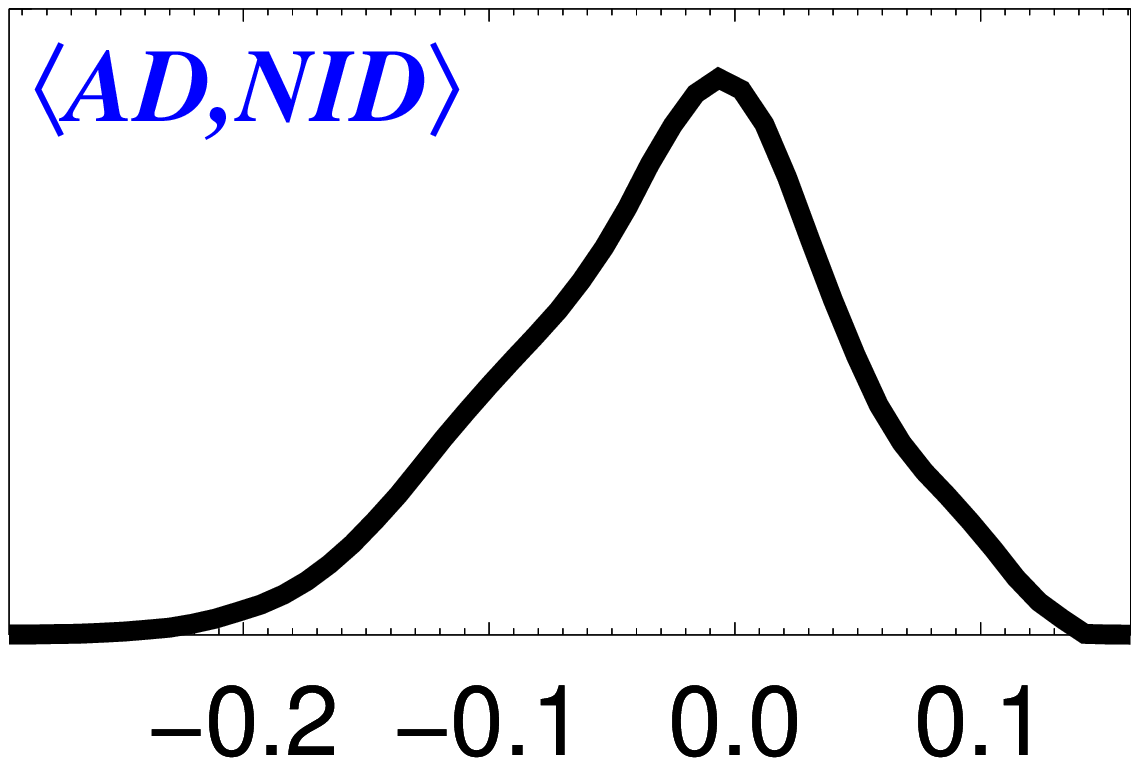}\hspace{-0.75cm}
\includegraphics[width=40mm]{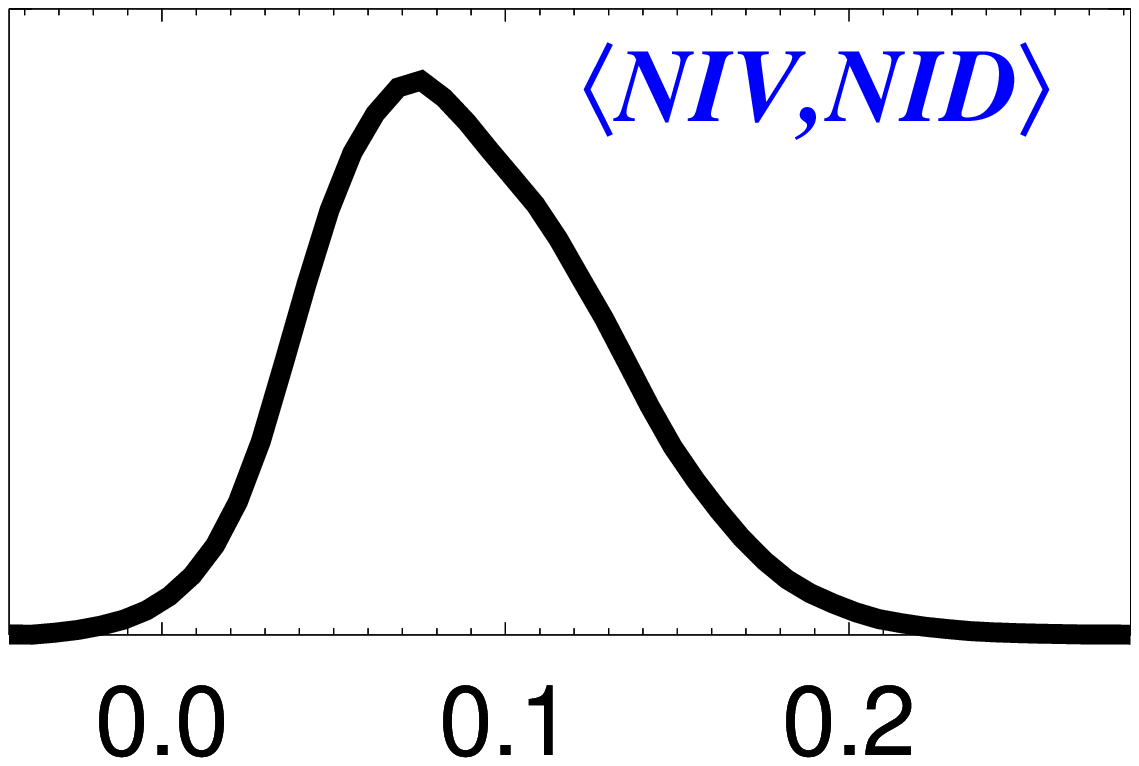}\hspace{-0.75cm}
\includegraphics[width=40mm]{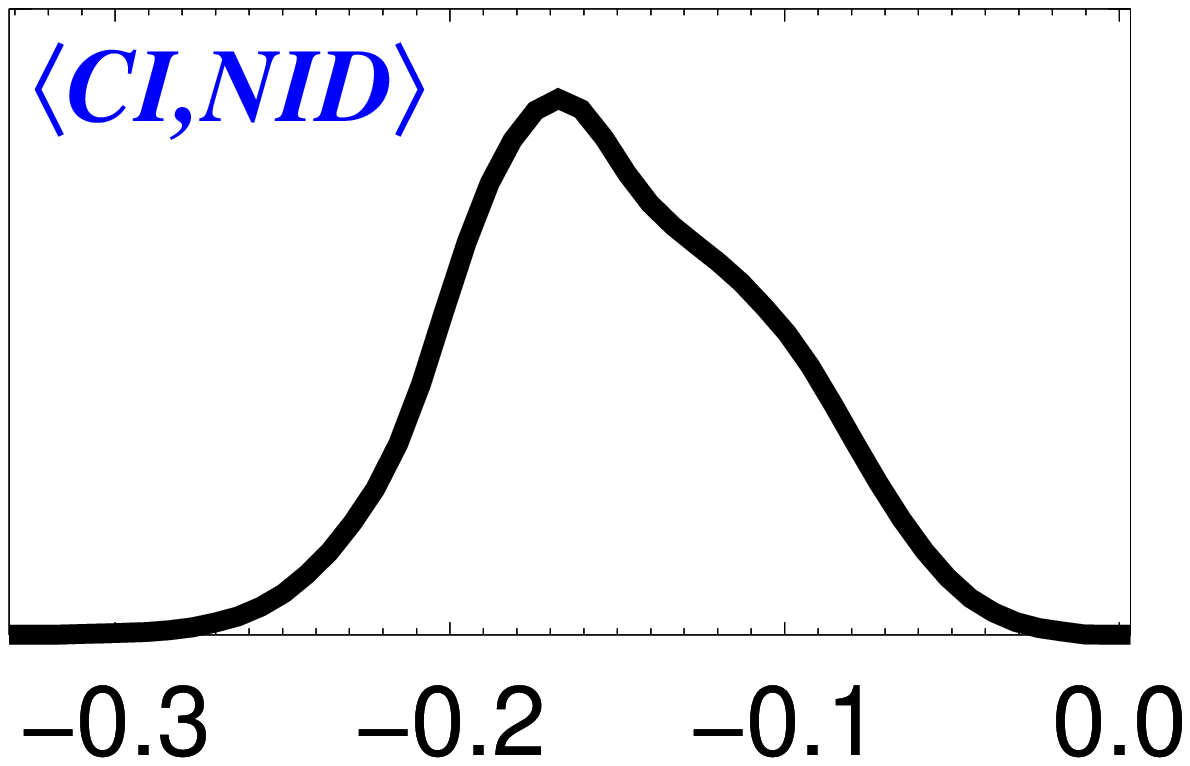}\hspace{-0.75cm}
\includegraphics[width=40mm]{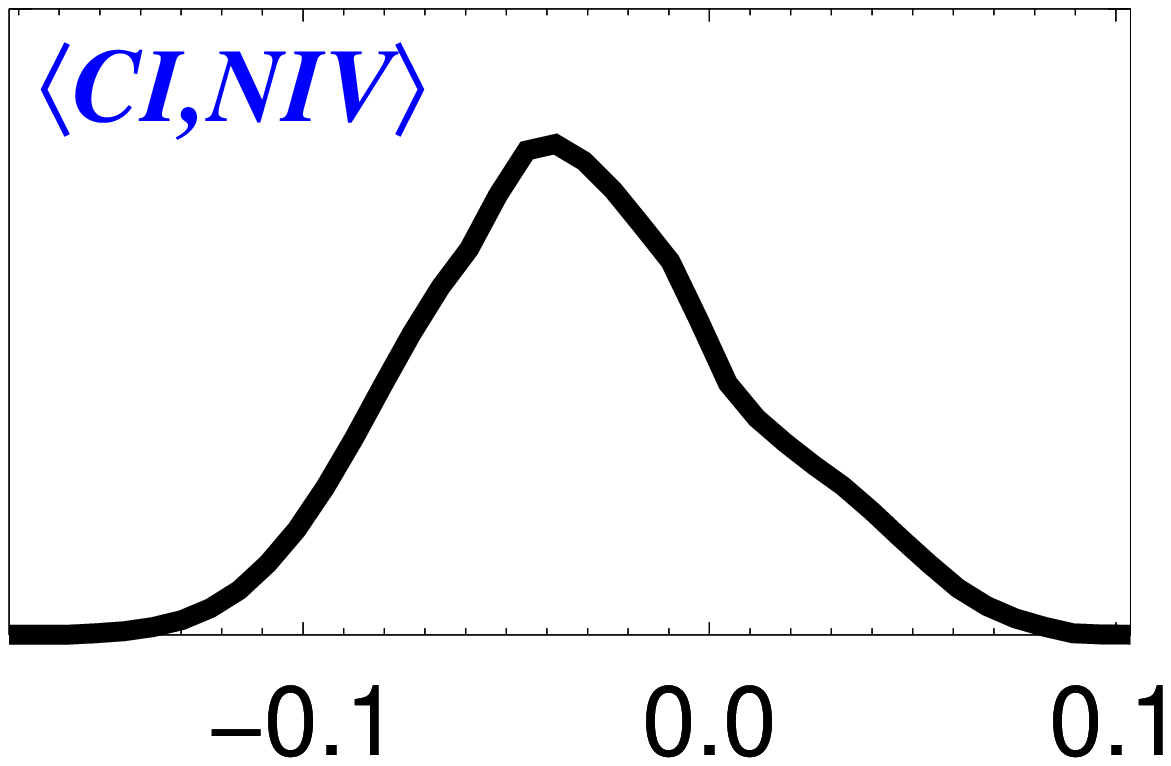}
\caption{ {\bf Admixture model}. Marginal posterior distributions for the six cosmological parameters and the auto-correlation and cross-correlation amplitudes in the admixture model. The solid line represents the admixture model constraints with the dashed line representing the adiabatic model constraints.}
 \label{fig:cosmoparms_AD_ISO_allmode}
\end{figure}

We now present cosmological constraints for the general admixture model with an adiabatic mode correlated with all three isocurvature modes. The marginalised posterior distributions of the parameters are shown in Figure \ref{fig:cosmoparms_AD_ISO_allmode} with the statistics and relative mode amplitudes given in Table \ref{table:allmodeISO}. There is a noticeable increase in the baryon density, $\Omega_{b}h^2$, although it is less than the lower limit found using WMAP three-year data \citep{Bean2006PhRvD_74f3503B} and WMAP first-year data \citep{Moodley2004PhRvD_70j3520M}. The baryon density is inconsistent with the adiabatic model value at approximately the $2\sigma$ level. All other cosmological parameters are consistent with the adiabatic model, though the admixture model prefers a slightly higher value of the cold dark matter density, and the constraint on the overall amplitude is much broader in this model. We find a permitted isocurvature fraction of $36 \pm 3 \%$, which is smaller than the $44\%$ and $60\%$ found using WMAP third-year \citep{Bean2006PhRvD_74f3503B} and first-year \citep{Moodley2004PhRvD_70j3520M} data respectively.\\
\begin{table}
\centering
\begin{tabular}{|c|c|c|c|}
\hline
\hline
Parameter			& mean $\pm$ $1\sigma$	& Parameter					& mean $\pm$ $1\sigma$	\\
\hline
$\Omega_{b}h^2$		& $0.025 \pm 0.001$		& $Z_{\left<NIV,NIV\right>}$	& $0.08 \pm 0.03$ \\
$\Omega_{c}h^2$		& $0.117  \pm 0.005$	& $Z_{\left<AD,CI\right>}$	& $0.03 \pm 0.08$ \\
$\Omega_{\Lambda}$	& $0.73  \pm  0.03$	 	& $Z_{\left<AD,NID\right>}$	& $-0.02 \pm 0.04$ \\
$\tau$				& $0.09  \pm  0.01$		& $Z_{\left<AD,NIV\right>}$	& $-0.01 \pm 0.06$ \\
$n_s$				& $0.99  \pm  0.02$		& $Z_{\left<CI,NID\right>}$	& $-0.22 \pm 0.04$ \\
$A_s$   			& $17.29   \pm  0.87$	& $Z_{\left<CI,NIV\right>}$	& $-0.10 \pm 0.02$ \\
$Z_{\left<AD,AD\right>}$	& $0.88 \pm 0.02$	& $Z_{\left<NID,NIV\right>}$	& $0.10 \pm 0.04$ \\
$Z_{\left<CI,CI\right>}$	& $0.17 \pm 0.04$	& $Z_{ISO}$	& $0.50 \pm 0.05$ \\
$Z_{\left<NID,NID\right>}$	& $0.19 \pm 0.05$	& $f_{ISO}$	& $0.36 \pm 0.03$ \\
\hline
\end{tabular}
\caption{ {\bf Admixture model}. Statistics derived from converged samples of the PMC algorithm for general admixture model parameters. The mean values and $1\sigma$ errors for each parameter are listed.}    
\label{table:allmodeISO}
\end{table}

\begin{figure}
    \begin{center}
       \subfigure{
           \includegraphics[width=60mm]{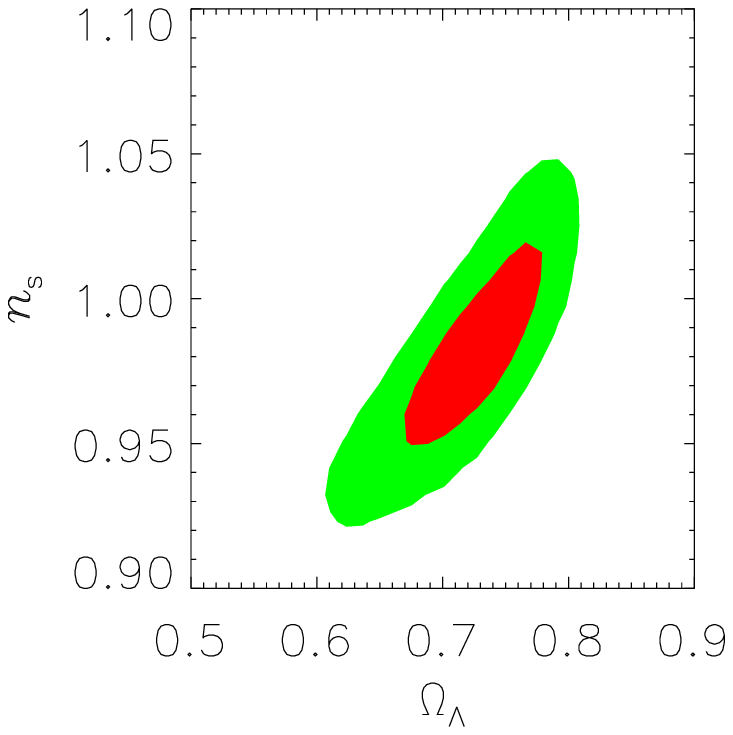}
       }\hspace{-0.15cm}%
       \subfigure{
          \includegraphics[width=60mm]{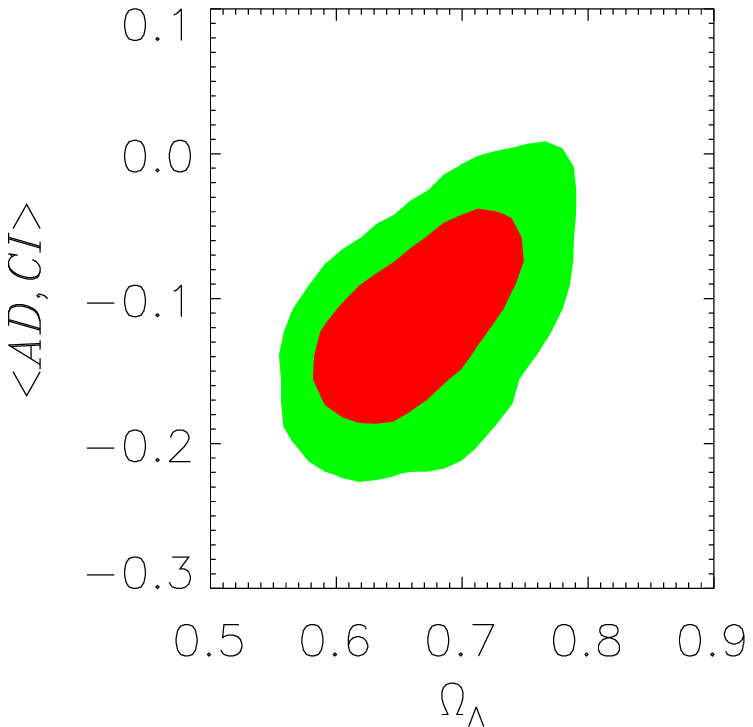}
       }\\ 
       \vspace{-0.5cm}
       \subfigure{
           \includegraphics[width=60mm]{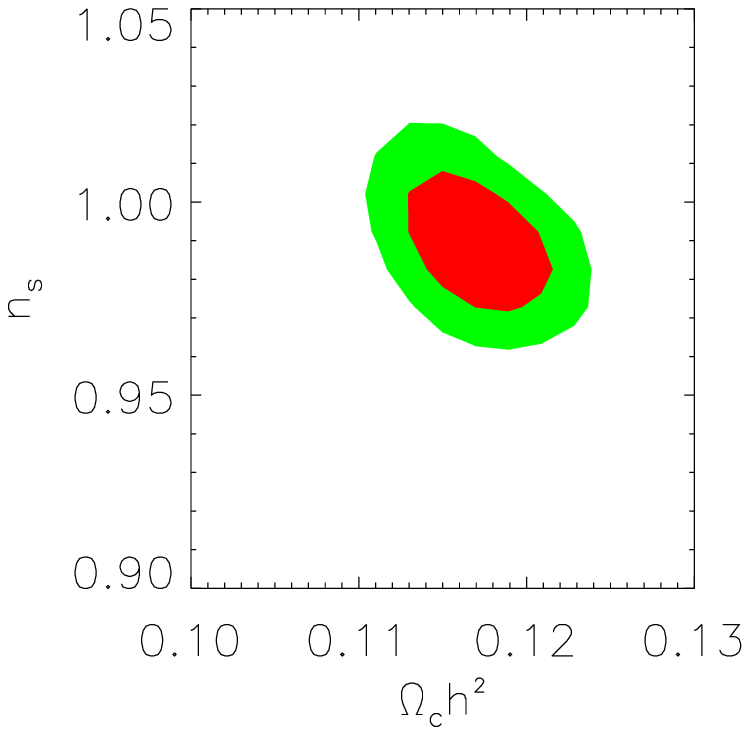}
       }\hspace{-0.15cm}%
       \subfigure{
           \includegraphics[width=60mm]{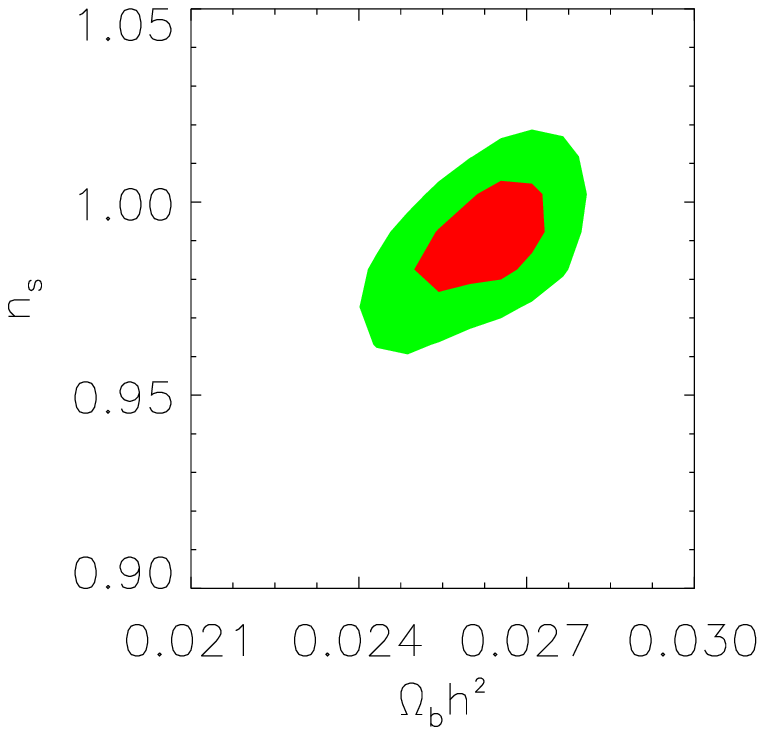}
       }
   \end{center}
   \caption{{\bf Admixture model}. Joint distributions between certain parameters are plotted to illustrate
the degeneracies present in the admixture model parameter space.  Confidence regions of $1\sigma$ and $2\sigma$ contours are shown in the green (lighter) and red (darker) colours respectively.}%
  \label{fig:allmodecontoursdegen}
\end{figure}

\noindent
There are strong degeneracies between parameters in the admixture model, as shown in Figure \ref{fig:allmodecontoursdegen}, which also shows the strong positive correlation of $\Omega_{\Lambda }$ and $n_s$. We use a principal component analysis to identify the main parameters contributing to the principal degeneracy. We find that these parameters are $\left<\mbox{AD,CI}\right>$, $\left< \mbox{AD,AD} \right>$, $\left<\mbox{NID,NID}\right>$, $\left<\mbox{CI,NID}\right>$, $\left<\mbox{AD,NID}\right>$ and $\Omega_{\Lambda}$, similar to the degeneracy identified using WMAP first-year data \citep{Moodley2004PhRvD_70j3520M}. These parameters collectively contribute to over $90\%$ of the degenerate direction. We also searched for other degenerate directions and found a second degeneracy in which the spectral index is strongly correlated with the baryon density and cold dark matter density.\\

\begin{figure}
\begin{center}
\includegraphics[width=150mm]{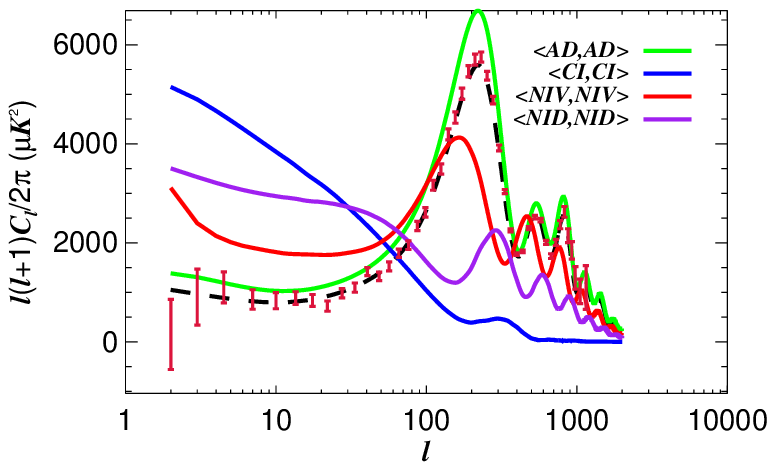}
\end{center}
\caption{{\bf Admixture model}. The angular power spectrum of CMB temperature anisotropies, $C_{\ell}$, for an admixture model (black-dashed curve) with large isocurvature fraction ($f_{ISO}=39\%$). The admixture model spectrum is a good fit to the WMAP nine-year data represented by the red error bars. To keep the plot uncluttered we omit the cross-correlation spectra that contribute to the admixture model, and only show the contributions from the auto-correlation modes.}
\label{fig:speccompallmode}
\end{figure}

\noindent
In Figure \ref{fig:speccompallmode}, we plot the CMB spectrum for a mixed model specified by ($\Omega_{b}h^2 , \Omega_{c}h^2 ,\Omega_{\Lambda } , \tau , n_s , A_s , f_{ISO}$) = (0.026, 0.125, 0.679, 0.099, 0.958, 18.22, 0.39) along with the individual auto-correlation mode contributions. This model has a high isocurvature fraction ($39\%$) and relative log-likelihood to the adiabatic model of $\frac{1}{2}\Delta\chi^{2} = \ln \mathcal{L}_{\text{AD} }-\ln \mathcal{L}_{\rm admix}=0.24$, where $\mathcal{L}_{\text{AD} }$ represents the likelihood of the best-fit pure adiabatic model and $\mathcal{L}_{\rm admix}$ the likelihood of the best-fit admixture model. An admixture model with a reasonably large isocurvature fraction is thus capable of producing CMB spectra that provide an equally good fit to the WMAP nine-year data. However, the extra degrees of freedom will downweight the admixture model in a Bayesian model selection framework, an investigation that we leave for a later study. 

\section{Conclusion}
\label{sec:conclusion}
In this paper we have established various methods for optimising the performance of the PMC algorithm. For the case of a Gaussian target distribution we found that the sampler is optimised with just two components and the optimal sample size scales quadratically with dimension. For more complex distributions, specifically banana shaped distributions, we found it necessary to increase the number of components. Bimodal distributions proved most challenging when peaks are separated by a large region of low probability.\\

We thereafter applied our simulation findings to constrain an admixture of adiabatic and isocurvature perturbations using CMB data from the WMAP nine-year release. We showed that the six parameters of the pure adiabatic model can be sampled with as little as two components and a sample size equivalent to the optimal sample size found for a Gaussian target distribution of six dimensions. A key strategy for the PMC algorithm that we inferred from this study is that it is worthwhile starting with a larger number of components initially and using a less conservative sample size. Based on the results from the first run, the sample size and number of components could be reduced in subsequent runs, thereby reducing the computational cost. \\

The parameter space of the admixture model that we studied is complex with significant degeneracies that require a large number of components to sample. Using the PMC algorithm on such a model resulted in significant degeneracies amongst the importance weights, which required us to implement the Non-linear Population Monte Carlo algorithm, specifically the soft clipping transformation.\\

The results for the admixture model indicate a smaller allowed isocurvature fraction compared to previous studies using earlier WMAP data releases. The covariance matrix for the admixture model derived in this paper using the WMAP nine-year dataset will serve as useful prior information for constraints on isocurvature perturbations with future datasets. Specifically we expect that the higher precision CMB temperature and polarisation power spectra measured by Planck will yield much stronger constraints on isocurvature perturbations. This study will be pursued in a future paper.

\acknowledgments

This effort is financially supported by the South African SKA Project (SKA SA) in the form of a PhD bursary to DM. DM sincerely thanks the SKA SA for funding this research and the PhD degree. The support and resources from the Centre for High Performance Computing (CHPC) in Cape Town are acknowledged. We acknowledge Martin Bucher for originally suggesting this study. 

\bibliographystyle{simon_bib_default}
\bibliography{PMC_paper}

\begin{thebibliography}{10}
\providecommand{\url}[1]{\texttt{#1}}
\providecommand{\urlprefix}{URL }
\providecommand{\eprint}[2][]{\url{#2}}

\bibitem{Runyan2003ApJS_149_265R}
M.~C. {Runyan} \textit{et~al.}, \apjs \textbf{149}, 265 (2003),
  \eprint{astro-ph/0303515}.

\bibitem{Kosowsky2003NewAR..47..939K}
A.~{Kosowsky}, \nar \textbf{47}, 939 (2003), \eprint{astro-ph/0402234}.

\bibitem{Tristram2004ASSL..301...97T}
M.~{Tristram} \& {Archeops Collaboration}, M.~{Plionis}, ed.,
  \textit{Astrophysics and Space Science Library}, vol. 301 of
  \textit{Astrophysics and Space Science Library}, 97 (2004),
  \eprint{astro-ph/0309349}.

\bibitem{Ade2014ApJ_792_62A}
P.~A.~R. {Ade} \textit{et~al.}, \apj \textbf{792}, 62 (2014),
  \eprint{1403.4302}.

\bibitem{MacTavish2006ApJ...647..799M}
C.~J. {MacTavish} \textit{et~al.}, \apj \textbf{647}, 799 (2006),
  \eprint{astro-ph/0507503}.

\bibitem{Padin2001ApJ...549L...1P}
S.~{Padin} \textit{et~al.}, \apjl \textbf{549}, L1 (2001),
  \eprint{astro-ph/0012211}.

\bibitem{Mather1990ApJ...354L..37M}
J.~C. {Mather} \textit{et~al.}, \apjl \textbf{354}, L37 (1990).

\bibitem{Rabii2006RScI...77g1101R}
B.~{Rabii} \textit{et~al.}, Review of Scientific Instruments \textbf{77},
  071101 (2006), \eprint{astro-ph/0309414}.

\bibitem{Bouchet2014arXiv1405_0439B}
F.~R. {Bouchet} \& {on behalf of the Planck collaboration for the results},
  ArXiv e-prints  (2014), \eprint{1405.0439}.

\bibitem{Miller2002ApJS..140..115M}
A.~{Miller} \textit{et~al.}, \apjs \textbf{140}, 115 (2002),
  \eprint{astro-ph/0108030}.

\bibitem{Ruhl2004SPIE.5498...11R}
J.~{Ruhl} \textit{et~al.}, \textit{{The South Pole Telescope}}, vol. 5498 of
  \textit{Society of Photo-Optical Instrumentation Engineers (SPIE) Conference
  Series}, 11--29 (2004), \eprint{astro-ph/0411122}.

\bibitem{Bennett2003ApJ_583_1B}
C.~L. {Bennett} \textit{et~al.}, \apj \textbf{583}, 1 (2003),
  \eprint{astro-ph/0301158}.

\bibitem{Barkats2003NewAR..47.1077B}
D.~{Barkats}, \nar \textbf{47}, 1077 (2003), \eprint{astro-ph/0306002}.

\bibitem{Leitch2002ApJ...568...28L}
E.~M. {Leitch} \textit{et~al.}, \apj \textbf{568}, 28 (2002),
  \eprint{astro-ph/0104488}.

\bibitem{Johnson2003NewAR..47.1067J}
B.~R. {Johnson} \textit{et~al.}, \nar \textbf{47}, 1067 (2003),
  \eprint{astro-ph/0308259}.

\bibitem{Wollack1997ApJ...476..440W}
E.~J. {Wollack} \textit{et~al.}, \apj \textbf{476}, 440 (1997),
  \eprint{astro-ph/9601196}.

\bibitem{Ade2008ApJ_674_22A}
P.~{Ade} \textit{et~al.}, \apj \textbf{674}, 22 (2008), \eprint{0705.2359}.

\bibitem{QUIET2012arXiv1207_5562Q}
{QUIET Collaboration} \textit{et~al.}, ArXiv e-prints  (2012),
  \eprint{1207.5562}.

\bibitem{Watson2003MNRAS.341.1057W}
R.~A. {Watson} \textit{et~al.}, \mnras \textbf{341}, 1057 (2003),
  \eprint{astro-ph/0205378}.

\bibitem{Planck2015arXiv150202114P}
{Planck Collaboration} \textit{et~al.}, ArXiv e-prints  (2015),
  \eprint{1502.02114}.

\bibitem{Planck2015arXiv150201589P}
{Planck Collaboration} \textit{et~al.}, ArXiv e-prints  (2015),
  \eprint{1502.01589}.

\bibitem{Bennett2013ApJS..208...20B}
C.~L. {Bennett} \textit{et~al.}, \apjs \textbf{208}, 20 (2013),
  \eprint{1212.5225}.

\bibitem{Bayes63}
M.~Bayes \& M.~Price, Philosophical Transactions \textbf{53}, 370 (1763),
  \eprint{http://rstl.royalsocietypublishing.org/content/53/370.full.pdf+html}.

\bibitem{Lindley2005}
D.~Lindley, Journal of the Royal Statistical Society: Series A (Statistics in
  Society) \textbf{168}, 259 (2005).

\bibitem{sivia2006data}
D.~Sivia \& J.~Skilling, \textit{Data Analysis: A Bayesian Tutorial}, Oxford
  science publications, OUP Oxford (2006).

\bibitem{robert2007bayesian}
C.~Robert, \textit{The Bayesian Choice: From Decision-Theoretic Foundations to
  Computational Implementation}, Springer texts in statistics, Springer (2007).

\bibitem{Skilling06}
J.~{Skilling}, R.~{Fischer}, R.~{Preuss} \& U.~V. {Toussaint}, eds.,
  \textit{American Institute of Physics Conference Series}, vol. 735 of
  \textit{American Institute of Physics Conference Series}, 395--405 (2004).

\bibitem{Mukherjee06}
P.~{Mukherjee}, D.~{Parkinson} \& A.~R. {Liddle}, Astrophysical Journal Letters
  \textbf{638}, L51 (2006), \eprint{arXiv:astro-ph/0508461}.

\bibitem{Parkinson06}
D.~{Parkinson}, P.~{Mukherjee} \& A.~R. {Liddle}, Physical review D
  \textbf{73}, 123523 (2006), \eprint{arXiv:astro-ph/0605003}.

\bibitem{Nikolic09}
B.~{Nikolic}, ArXiv e-prints  (2009), \eprint{arXiv:0912.2317}.

\bibitem{Valiviita09}
J.~{V{\"a}liviita} \& T.~{Giannantonio}, Physical review D \textbf{80}, 123516
  (2009), \eprint{arXiv:0909.5190}.

\bibitem{Feroz09}
F.~{Feroz}, M.~P. {Hobson} \& M.~{Bridges}, Monthly Notices of the Royal
  Astrophysical Society \textbf{398}, 1601 (2009), \eprint{arXiv:0809.3437}.

\bibitem{Metropolis1953}
N.~{Metropolis} \textit{et~al.}, Journal of Chemical Physics \textbf{21}, 1087
  (1953).

\bibitem{2002PhRvD..66j3511L}
A.~{Lewis} \& S.~{Bridle}, Physical Review D \textbf{66}, 103511 (2002),
  \eprint{arXiv:astro-ph/0205436}.

\bibitem{Oh92}
M.~S. {Oh} \& J.~O. {Berger}, Journal of Statistical Computation and Simulation
  \textbf{41}, 143 (1992).

\bibitem{Cappe2004}
O.~Capp{\'e} \textit{et~al.}, Journal of Computational and Graphical Statistics
  \textbf{13}, pp. 907 (2004).

\bibitem{Wraith09}
D.~{Wraith} \textit{et~al.}, Physical Review D \textbf{80}, 023507 (2009),
  \eprint{0903.0837}.

\bibitem{Dietrich2010MNRAS}
J.~P. {Dietrich} \& J.~{Hartlap}, Monthly Notices of the Royal Astrophysical
  Society \textbf{402}, 1049 (2010), \eprint{0906.3512}.

\bibitem{Kilbinger10}
M.~{Kilbinger} \textit{et~al.}, Monthly Notices of the Royal Astrophysical
  Society \textbf{405}, 2381 (2010), \eprint{0912.1614}.

\bibitem{Kilbinger2011CosmoPMC}
M.~{Kilbinger} \textit{et~al.}, ArXiv e-prints  (2011), \eprint{1101.0950}.

\bibitem{Dunkley04}
J.~{Dunkley} \textit{et~al.}, Monthly Notices of the Royal Astrophysical
  Society \textbf{356}, 925 (2005), \eprint{arXiv:astro-ph/0405462}.

\bibitem{Vita2012ApJ_753_151V}
J.~{V{\"a}liviita} \textit{et~al.}, \apj \textbf{753}, 151 (2012),
  \eprint{1202.2852}.

\bibitem{Savelainen2013PhRvD_88f3010S}
M.~{Savelainen} \textit{et~al.}, Physical Review D \textbf{88}, 063010 (2013),
  \eprint{1307.4398}.

\bibitem{Planck2015arXiv150702704P}
{Planck Collaboration} \textit{et~al.}, ArXiv e-prints  (2015),
  \eprint{1507.02704}.

\bibitem{Kullback1951}
S.~Kullback \& R.~A. Leibler, The Annals of Mathematical Statistics
  \textbf{22}, pp. 79 (1951).

\bibitem{Cappe07}
O.~{Capp{\'e}} \textit{et~al.}, ArXiv e-prints  (2007), \eprint{0710.4242}.

\bibitem{Chopin2002}
N.~Chopin, Biometrika \textbf{89}, pp. 539 (2002).

\bibitem{Hinshaw2013}
G.~Hinshaw \textit{et~al.}, The Astrophysical Journal Supplement Series
  \textbf{208}, 19 (2013).

\bibitem{Bennet2013}
C.~L. Bennett \textit{et~al.}, The Astrophysical Journal Supplement Series
  \textbf{208}, 20 (2013).

\bibitem{Larson2011}
D.~Larson \textit{et~al.}, The Astrophysical Journal Supplement Series
  \textbf{192}, 16 (2011).

\bibitem{Dunkley2009}
J.~Dunkley \textit{et~al.}, The Astrophysical Journal Supplement Series
  \textbf{180}, 306 (2009).

\bibitem{Hinshaw2003}
G.~Hinshaw \textit{et~al.}, The Astrophysical Journal Supplement Series
  \textbf{148}, 135 (2003).

\bibitem{Verde2003}
L.~Verde \textit{et~al.}, The Astrophysical Journal Supplement Series
  \textbf{148}, 195 (2003).

\bibitem{Lewis2000}
A.~Lewis \& S.~Bridle, Phys. Rev. D \textbf{66}, 103511 (2002).

\bibitem{Moodley2004PhRvD_70j3520M}
K.~{Moodley} \textit{et~al.}, Physical Review D \textbf{70}, 103520 (2004),
  \eprint{astro-ph/0407304}.

\bibitem{Kasanda_Moodley2014}
S.~{Muya Kasanda} \& K.~{Moodley}, \jcap \textbf{12}, 041 (2014),
  \eprint{1409.6869}.

\bibitem{Koblents2012arXiv1208_5600K}
E.~{Koblents} \& J.~{M{\'{\i}}guez}, ArXiv e-prints  (2012),
  \eprint{1208.5600}.

\bibitem{Bean2006PhRvD_74f3503B}
R.~{Bean}, J.~{Dunkley} \& E.~{Pierpaoli}, \prd \textbf{74}, 063503 (2006),
  \eprint{astro-ph/0606685}.

\end{thebibliography}




\end{document}